\documentclass[prd,aps,preprint,groupedaddress,superscriptaddress,nofootinbib]{revtex4}
\usepackage{graphicx}
\usepackage{amsmath,amssymb,bm}
\usepackage{hyperref}
\usepackage{autobreak,mathtools}
\usepackage[dvipsnames, usenames]{xcolor}

\definecolor{p-r}{RGB}{171, 40, 52}
\hypersetup{colorlinks=true, citecolor=p-r, linkcolor=p-r,urlcolor=p-r}

\definecolor{changes}{RGB}{169.2, 155.0, 0.5}

\usepackage{ulem}

\allowdisplaybreaks[1]

\begin{document}

\preprint{YITP-25-185, RESCEU-28/25, IPMU25-0053}

\title{Macroscopic backreaction of the trace anomaly on classical vacuum backgrounds}
\author{Ra\'ul Carballo-Rubio}
\affiliation{Instituto de Astrof\'isica de Andaluc\'ia (IAA-CSIC),
Glorieta de la Astronom\'ia, 18008 Granada, Spain}
\affiliation{Center of Gravity, Niels Bohr Institute, Blegdamsvej 17, 2100 Copenhagen, Denmark}
\author{Francesco Di Filippo}
\affiliation{Institut f\"ur Theoretische Physik, Max-von-Laue-Str. 1, 60438 Frankfurt, Germany}
\author{Shinji Mukohyama}
\affiliation{Center for Gravitational Physics and Quantum Information, Yukawa Institute for Theoretical Physics, Kyoto University, Kyoto 606-8502, Japan}
\affiliation{Research Center for the Early Universe (RESCEU), Graduate School of Science, The University of Tokyo, Hongo 7-3-1, Bunkyo-ku, Tokyo 113-0033, Japan}
\affiliation{Kavli Institute for the Physics and Mathematics of the Universe (WPI),
The University of Tokyo Institutes for Advanced Study,
The University of Tokyo, Kashiwa, Chiba 277-8583, Japan}
\author{Kazumasa Okabayashi}
\affiliation{Center for Gravitational Physics and Quantum Information, Yukawa Institute for Theoretical Physics, Kyoto University, Kyoto 606-8502, Japan}

\begin{abstract}
We study the backreaction of quantum fields in the Boulware vacuum state on the Schwarzschild geometry, using the Riegert--Mottola--Vaulin renormalized stress-energy tensor derived from the conformal anomaly. An order-reduction procedure is applied to first order, paying special attention to the conservation of the resulting stress-energy tensor. The results obtained in these different situations are compared between them, and also to recent works in the literature using other approximations for the renormalized stress-energy tensor.
\end{abstract}

\maketitle

\section{Introduction}

Studies of quantum effects in the presence of gravitational fields play a prominent role in theoretical gravitational physics. These studies can be broadly divided into the study of quantum fields in curved backgrounds~\cite{Birrell:1982ix,Mukhanov:2007zz}, and the study of the quantum structure of the gravitational interaction itself~\cite{Rovelli:2004tv,Oriti:2009zz}. These two aspects are nevertheless related, as it is expected that any quantum theory of gravity will have a semiclassical regime described by some form of the semiclassical Einstein equations~\cite{Wald:1984rg,Singh:1989ct}. These semiclassical Einstein equations describe the backreaction of quantum fields on the curved spacetime metric. For instance, the existence of this semiclassical regime is an implicit assumption in the paradigm of black hole evaporation due to the emission of Hawking radiation~\cite{Hawking:1974rv,Hawking:1975vcx,Unruh:1980cg,Parentani:1994ij}, which is a pillar of black hole physics.

A growing body of literature examines static and spherically symmetric solutions of semiclassical equations, in particular describing stellar interiors~\cite{Carballo-Rubio:2017tlh,Ho:2017vgi,Arrechea:2021pvg,Arrechea:2021xkp,Arrechea:2023oax} and solutions devoid of matter that can be understood as their environments~\cite{Fabbri:2005zn,Ho:2017joh,Berthiere:2017tms,Ho:2018jkm,Arrechea:2019jgx,Arrechea:2021ldl,Arrechea:2022dvy,Beltran-Palau:2022nec}. These works present cumulative evidence of the existence of new forms of stellar equilibrium that can represent intermediate states between neutron stars and black holes~\cite{Carballo-Rubio:2017tlh,Arrechea:2021xkp,Arrechea:2023wgy}. These theoretical developments may provide a solid theoretical foundation for phenomenological studies of black hole mimickers~\cite{Cardoso:2017cqb,Carballo-Rubio:2018vin,Barausse:2018vdb,Carballo-Rubio:2018jzw,Cardoso:2019rvt,Cardoso:2019nis,Maggio:2020jml,Maggio:2021uge,Carballo-Rubio:2022imz,Carballo-Rubio:2022aed,Maggio:2022nme,Carballo-Rubio:2023fjj,Barcelo2011_v2,Harada:2018zfg,Kokubu:2019jdx,Okabayashi:2021qfg,Nakao_2022,Nakao_2023,Numajiri:2024qgh}, a topic that is receiving increasing attention~\cite{Carballo-Rubio:2025fnc,Bambi:2025wjx} as new datasets from gravitational-wave~\cite{PhysRevLett.116.061102,LIGOScientific:2017bnn,LIGOScientific:2017vwq,LIGOScientific:2019fpa,LIGOScientific:2019zcs,LIGOScientific:2020tif,LIGOScientific:2021nrg,LIGOScientific:2021sio} and electromagnetic~\cite{EventHorizonTelescope:2019dse,EventHorizonTelescopeI,EventHorizonTelescopeII,EventHorizonTelescopeIII,EventHorizonTelescopeIV,EventHorizonTelescopeV,EventHorizonTelescopeVI} observations are being released.

In this paper, we use an explicit analytical approximation to the renormalized stress-energy tensor of quantum fields, the Riegert--Mottola--Vaulin Renormalized Stress-Energy Tensor (RMV-RSET), as the source of the gravitational field~\cite{Riegert1984PhLB..134...56R,Mottola:2006ew}. This allows us to study how quantum effects, as encoded in the RMV-RSET, modify the structure of the classical Schwarzschild solution.\footnote{In \cite{Numajiri:2024qgh}, on the other hand, impacts of
the boundary condition at the center of a static horizonless regular
spacetime have been studied in the context of the RMV-RSET.} While the RMV-RSET fails to capture Weyl-invariant contributions to the renormalized stress-energy tensor, as pointed out by Deser ~\cite{Deser:1996qa,Deser:1999zv} (see also~\cite{Barvinsky:2023exr}), it has been used before for the study of infrared modifications of general relativity in other situations~\cite{Mottola:2006ew,Mottola:2016mpl,Mottola:2022tcn}, and provides a series of advantages with respect to the approximations previously used in the literature to study the backreaction of quantum fields on classical vacuum backgrounds~\cite{Fabbri:2005zn,Ho:2017joh,Berthiere:2017tms,Ho:2018jkm,Arrechea:2019jgx,Arrechea:2021ldl,Arrechea:2022dvy,Beltran-Palau:2022nec}. These advantages, as well as the similarities and differences with respect to the results reported in these previous works, will be discussed below.

This paper is organized as follows. In Sec.~\ref{sec:RMV-RSET}, we introduce the RMV-RSET and briefly review its structure and properties as reported in previous papers, with the aim of making the discussion self-contained. In Sec.~\ref{sec:semi_eqs}, we introduce the semiclassical equations that we will be solving and discuss the concept of order reduction that will be central to our analysis. In Sec.~\ref{sec:order_red}, we present the order reduction of the RMV-RSET to first order and our numerical results, together with a comparison with previous results in the literature. Sec.~\ref{sec:conclusions} is devoted to a summary of the paper and discussions. We also provide three appendices to show some technical details.

\section{The Riegert--Mottola--Vaulin Renormalized Stress-Energy Tensor \label{sec:RMV-RSET}}

\subsection{General spacetimes}

The RMV-RSET, $\langle\hat{T}_{ab}\rangle$ in the following, is obtained by introducing auxiliary fields $\varphi$ and $\psi$ that allow expressing the non-local action leading to the conformal anomaly in a local form with fourth-order derivatives. To the best of our knowledge, this form of the RSET was first introduced by Riegert~\cite{Riegert1984PhLB..134...56R}, later used to reconstruct different quantum vacuum states on black hole backgrounds in~\cite{Balbinot:1999ri,Balbinot:1999vg}, and more recently stressed to be a source of infrared modifications to the Einstein equations by Mottola and Vaulin~\cite{Mottola:2006ew}, which is the  approach followed here as well (see also~\cite{Giannotti:2008cv,Mottola:2010gp,Mottola:2016mpl,Mottola:2022tcn}). While the RMV-RSET represents an approximation which is known to drop Weyl-invariant contributions to the renormalized stress-energy tensor ~\cite{Deser:1996qa,Deser:1999zv}, there are general arguments pointing that the anomaly-induced terms retained in the approximation dominate over
Weyl-invariant terms in the near-horizon region~\cite{Mazur:2001aa}, consistently with results obtained in the Boulware state~\cite{Mottola:2025fhl}. There have also been proposals to obtain the expression for the trace anomaly with second-order equations of motion~\cite{Gabadadze:2023quw}. Here we deal with this problem differently, following an order-reduction procedure first introduced in~\cite{Simon:1990jn,Parker:1993dk}.

We summarize the main properties of the RMV-RSET as described in~\cite{Mottola:2006ew}, with the goal  of solving the semiclassical Einstein field equations. The RMV-RSET provides an analytical approximation to the RSET of conformal matter fields of arbitrary spin ($\leq 1$) in an arbitrary curved spacetime, given by
\begin{equation}\label{eq:rmv_def}
\langle \hat{T}_{ab}\rangle = b' E_{ab} + b F_{ab}\,,
\end{equation}
where $b$ and $b'$ are numerical coefficients,
\begin{equation}
b= \frac{\hbar}{120 (4 \pi)^2}\, (N_S + 6 N_F + 12 N_V)\,,\qquad b' = -\frac{\hbar}{360 (4 \pi)^2}\, (N_S + 11 N_F + 62 N_V)\,,
\end{equation}
with $N_S$ the number of spin-0 fields, $N_F$ the number of spin-1/2 Dirac fields, and $N_V$ the number of spin-1 fields (these numerical coefficients come directly from the conformal anomaly, as described, e.g., in~\cite{Birrell:1982ix}).

The tensors $E_{ab}$ and $F_{ab}$ are defined as follows:
\begin{align}\label{eq:Edef}
E_{ab} =& -2\nabla_{(a}\varphi\nabla_{b)}\square \varphi
+ 2\nabla^c \left(\nabla_c \varphi\nabla_a\nabla_b\varphi\right)
- \frac{2}{3}\nabla_a\nabla_b\left(\nabla_c \varphi
\nabla^c\varphi\right)\nonumber\\
&+ \frac{2}{3}R_{ab}\nabla_c \varphi\nabla^c \varphi
- 4R^c_{\ (a}\nabla_{b)} \varphi\nabla_c \varphi
+ \frac{2}{3}R\nabla_a \varphi\nabla_b \varphi\nonumber\\
& + \frac{1}{6}g_{ab}\left[-3(\square\varphi)^2
+ \square \left(\nabla_c\varphi\nabla^c\varphi\right)
+ 2\left( 3R^{cd} - R g^{cd} \right) \nabla_c \varphi\nabla_d
\varphi\right]\nonumber\\
&- \frac{2}{3}\nabla_a\nabla_b\square\varphi
- 4C_{a\ b}^{\ c\ d}\nabla_c \nabla_d \varphi
- 4R_{(a}^c \nabla_{b)} \nabla_c\varphi
+ \frac{8}{3}R_{ab}\square\varphi
+ \frac{4}{3}R\nabla_a\nabla_b\varphi \nonumber\\
&- \frac{2}{3} \nabla_{(a}R \nabla_{b)}\varphi
+ \frac{1}{3}g_{ab}\left(2\square ^2 \varphi
+ 6R^{cd}\nabla_c\nabla_d\varphi
- 4R\square \varphi
+ \nabla^c R\nabla_c\varphi\right)\,,
\end{align}
and
\begin{align}\label{eq:Fdef}
 F_{ab} = &-2\nabla_{(a}\varphi\nabla_{b)} \square \psi
-2\nabla_{(a}\psi\nabla_{b)} \square \varphi
+ 2\nabla^c \left(\nabla_c \varphi\nabla_a\nabla_b\psi
 + \nabla_c \psi\nabla_a\nabla_b\varphi\right)
\nonumber\\
&- \frac{4}{3}\nabla_a\nabla_b\left(\nabla_c \varphi\nabla^c\psi\right)
+ \frac{4}{3}R_{ab}\nabla_c \varphi\nabla^c \psi
- 4R^c_{\ (a}\left(\nabla_{b)} \varphi\nabla_c \psi
+ \nabla_{b)} \psi\nabla_c \varphi\right)\nonumber\\
&+ \frac{4}{3}R\nabla_{(a} \varphi\nabla_{b)} \psi
+ \frac{1}{3}g_{ab}\left[-3\square\varphi\square\psi
+ \square \left(\nabla_c\varphi\nabla^c\psi\right)+2\left( 3R^{cd} - R g^{cd} \right)\nabla_c
\varphi\nabla_d \psi\right]\nonumber\\
&- 4\nabla_c\nabla_d\left( C_{(a\ b)}^{\ \ c\ \ d}
\varphi \right)  - 2C_{a\ b}^{\ c\ d} R_{cd} \varphi- \frac{2}{3}\nabla_a\nabla_b \square \psi
- 4C_{a\ b}^{\ c\ d}\nabla_c \nabla_d \psi\nonumber\\
 &- 4R_{(a}^c \nabla_{b)} \nabla_c\psi
+ \frac{8}{3}R_{ab}\square \psi
+ \frac{4}{3}R\nabla_a\nabla_b\psi - \frac{2}{3}\nabla_{(a}R \nabla_{b)}\psi \nonumber\\
&+
\frac{1}{3}g_{ab}\left(2\square^2 \psi + 6R^{cd}\nabla_c\nabla_d\psi
- 4R\square\psi +\nabla^c R\nabla_c\psi\right)\,.
\end{align}
Both quantities above are independently conserved. The auxiliary fields $\varphi$ and $\psi$ are required to satisfy the fourth-order equations
\begin{equation}
2 \Delta_4 \varphi = {^*R}_{abcd}{^*R}^{abcd} - \frac{2}{3} \square R\,,\quad
2 \Delta_4 \psi = C_{abcd}C^{abcd}\,,\label{eq:phipsieqs}
\end{equation}
so that the trace of the RSET~\eqref{eq:rmv_def} becomes
\begin{equation}
 g^{ab}\langle \hat{T}_{ab}\rangle = b' \left( {^*R}_{abcd}{^*R}^{abcd} - \frac{2}{3} \square R \right) + b\, C_{abcd}C^{abcd}\,,
\end{equation}
and we are using the definitions of the curvature invariants
\begin{align}\label{eq:cinvs}
{^*R}_{abcd}{^*R}^{abcd}&=R_{abcd}R^{abcd}-4R_{ab}R^{ab} + R^2\,,\nonumber\\
C_{abcd}C^{abcd} &= R_{abcd}R^{abcd}
-2 R_{ab}R^{ab}  + \frac{R^2}{3}\,,
\end{align}
as well as the derivative operator $\Delta_4$,
\begin{equation}\label{eq:delta4def}
\Delta_4=  \square^2 + 2 R^{ab}\nabla_a\nabla_b - \frac{2}{3} R \square +
\frac{1}{3} (\nabla^a R)\nabla_a \,.
\end{equation}

The expression for $\langle\hat{T}_{ab}\rangle$ above is completely determined once a background $g_{ab}$ is chosen and the fourth-order equations for $\varphi$ and $\psi$ are solved. While our aim in this paper is to solve self-consistently the Einstein equations with $\langle\hat{T}_{ab}\rangle$ acting as the source, it is useful to discuss first the form of $\langle\hat{T}_{ab}\rangle$ for specific fixed backgrounds, in particular the Schwarzschild spacetime. Relevant literature on the subject includes~\cite{Balbinot:1999ri,Balbinot:1999vg,Mottola:2006ew,Anderson:2007eu}. The same procedure applied below could also be applied to expressions that include Weyl-invariant contributions to the renormalized stress-energy tensor~\cite{Deser:1996qa,Deser:1999zv} (see also~\cite{Barvinsky:2023exr}) which, while being out of the scope of this paper, would be an interesting extension.

\subsection{Schwarzschild spacetime}

In this section, we will particularize the expressions above to the Schwarzschild metric
\begin{equation}
    \text{d}s^2=-f(r)\text{d}t^2+\frac{\text{d}r^2}{f(r)}+r^2\text{d}\Omega^2\,,
\end{equation}
with $\text{d}\Omega^2$ the line element on the unit 2-sphere and
\begin{equation}
f(r)=1-\frac{2M}{r}\,.
\end{equation}
We will always be using units in which $c=G=1$. The Schwarzschild metric is Ricci-flat, which implies that both curvature invariants in Eq.~\eqref{eq:cinvs} take the same value:
\begin{align}
{^*R}_{abcd}{^*R}^{abcd}=C_{abcd}C^{abcd}=\frac{48 M^2}{r^6}\,,
\end{align}
 while the differential operator defined in Eq.~\eqref{eq:delta4def} is reduced to
\begin{equation}\label{eq:delta4sch}
\Delta_4=  \square^2\,.
\end{equation}
Thus, the equations for $\varphi$ and $\psi$ in Eq.~\eqref{eq:phipsieqs} are the same, so we just focus on:
\begin{equation}
\square^2\varphi(r,t)=\frac{24M^2}{r^6}\,.
\end{equation}
A solution of this equation is given by the integral of the following expression~\cite{Balbinot:1999ri,Balbinot:1999vg,Mottola:2006ew}:
\begin{align}\label{eq:sol-phi-Sch}
\frac{\text{d}\varphi_0(r)}{\text{d}r}&=\frac{q-2}{6M}\,\left(\frac{r}{2M} + 1 + \frac{2M}{r}\right)
\ln \left(1 - \frac{2M}{r}\right) - \frac{q}{6r}\left[\frac{4M}{r-2M}
\ln \left(\frac{r}{2M}\right) + \frac{r}{2M} + 3 \right] \nonumber\\
&- \frac{1}{3M} - \frac{1}{r} + \frac{2M c_{_H}}{r (r-2M)}
+ \frac{c_\infty}{2M}\, \left(\frac{r}{2M} + 1 + \frac{2M}{r}\right)\,.
\end{align}
We are using the same notation as in~\cite{Mottola:2006ew} for the constants of integration
$c_{_H}$, $c_\infty$, and $q$. Note that the integration constant that results from integrating the equation above is irrelevant, as the RMV-RSET on a Ricci-flat background depends only on derivatives of $\varphi$. There is, however, an additional constant coming from the following time dependence that keeps the RMV-RSET time-independent:
\begin{equation}\label{eq:sol-phi-Sch_complete}
\varphi (r,t) = \varphi_0(r) + \frac{p}{2M}\,t\,.
\end{equation}
Different values of these integration constants select different quantum states, as discussed in detail in~\cite{Balbinot:1999ri,Balbinot:1999vg,Mottola:2006ew,Mottola:2010gp}. The integration constants of $\psi$ are, in general, different from those of $\varphi$ and are denoted as $d_{\rm H}$, $d_{\infty}$, $q'$, and $p'$ instead of $c_{\rm H}$, $c_{\infty}$, $q$, and $p$ in Eqs.~\eqref{eq:sol-phi-Sch} and~\eqref{eq:sol-phi-Sch_complete}.

From the asymptotic form of Eq.~\eqref{eq:sol-phi-Sch}, $\text{d}\varphi/\text{d}r$ and $\text{d}\psi/\text{d}r$ vanish in the $r\rightarrow\infty$ limit if and only if
\begin{align}\label{eq:cd-infty}
    c_{\infty}&=d_{\infty}=0, \quad \text{and} \quad q=q'=0\,.
\end{align}
To have a static ansatz, we also impose that
\begin{equation}\label{eq:p_pprime}
    p=p'=0\,.
\end{equation}
Following~\cite{Mottola:2006ew}, the remaining free parameters $c_{\rm H}$ and $d_{\rm H}$ are determined such that the RMV-RSET provides a good fit of the stress-energy tensor for a conformally coupled scalar field in the Boulware state~\cite{Mottola:2006ew}:
\begin{equation}\label{eq:ch_dh}
    c_{\rm H}=-\frac{7}{20},\qquad d_{\rm H}=\frac{55}{84}\,.
\end{equation}
We will use these values in the rest of the paper.

\section{Semiclassical Einstein equations and order reduction \label{sec:semi_eqs}}

In this section, we introduce the semiclassical Einstein equations and describe the procedure of order reduction as first implemented in~\cite{Simon:1990jn,Parker:1993dk}, giving a different interpretation to it along the lines of~\cite{Arrechea:2022dvy}. The starting point is the set of differential equations
\begin{equation}\label{eq:mateqs}
G_{ab}=8\pi\left(T_{ab}+\langle\hat{T}_{ab}\rangle\right)\,,
\end{equation}
where $G_{ab}$ is the Einstein tensor, $T_{ab}$ is a classical source , and $\langle \hat{T}_{ab}\rangle$ is the RMV-RSET. Note that the latter cannot be written as a local combination of curvature tensors, thus avoiding any contradiction with Lovelock's theorem~\cite{Lovelock:1971yv,Lovelock:1972vz}. Also, we are ignoring the effect of higher-order curvature corrections that may arise due to renormalization. From a mathematical perspective, Eq.~\eqref{eq:mateqs} provides a self-consistent set of differential equations that incorporates the backreaction of quantum vacuum fluctuations onto the spacetime geometry. The geometries we will be analyzing can be written without loss of generality as
\begin{equation}\label{eq:sssansatz}
    \text{d}s^2=-f(r)\text{d}t^2+h(r)\text{d}r^2+r^2\text{d}\Omega^2\,.
\end{equation}
We will assume asymptotically flat conditions so that, if $M$ is the Arnowitt--Deser--Misner (ADM) mass~\cite{Misner1964,Hernandez1966}, for $r\gg M$ we have:
\begin{equation}\label{eq:boundary_cond}
f(r)=\frac{1}{h(r)}+\mathcal{O}\left(\frac{M^2}{r^2}\right)=1-\frac{2M}{r}+\mathcal{O}\left(\frac{M^2}{r^2}\right)\,.
\end{equation}

The $\langle\hat{T}_{ab}\rangle$ term in Eq.~\eqref{eq:mateqs} is at least $\hbar$, which allows the implementation of an order-reduction procedure. This procedure can be understood as a way to perturbatively analyze the solutions to the system above, but also as a way to generate an inequivalent but simpler set of differential equations that still captures some of the main physical ingredients of the problem. Let us summarize the procedure described in~\cite{Simon:1990jn,Parker:1993dk}, performing an expansion in $\hbar$ in Eq.~\eqref{eq:mateqs} so that, at the lowest order, we can write
\begin{equation}
 G_{ab}+\mathcal{O}(\hbar)=8\pi T_{ab}+\mathcal{O}(\hbar)\,.
\end{equation}
This equation can be solved to obtain the Ricci tensor $R_{ab}$ in terms of $T_{ab}$.

In this paper, we will focus on vacuum situations, meaning that the classical source is absent. Hence, in practice we will be dealing with ``vacuum'' semiclassical equations, in the sense that the expectation value of the stress-energy tensor on the right-hand side is the only source, and also that the expectation value is taken on a vacuum state.

As $\langle\hat{T}_{ab}\rangle$ is not written as a local combination of curvature tensors, additional considerations are needed to implement the order-reduction procedure of this quantity. Restricting our consideration to the vacuum situation ($T_{ab}=0$) and the ansatz in Eq.~\eqref{eq:sssansatz}, we can obtain explicitly all derivatives of the functions $f(r)$ and $h(r)$ as the following polynomials~\cite{Arrechea:2022dvy}:
\begin{align}\label{eq:derrels}
f^{(n)}&=(-1)^{n+1}\frac{n!f}{r^n}(h-1)+\mathcal{O}(\hbar)\,,\nonumber\\
h^{(n)}&=(-1)^{n}\frac{n!h^n}{r^n}(h-1)+\mathcal{O}(\hbar)\,.
\end{align}
In~\cite{Arrechea:2022dvy}, these expressions were used to reduce the Anderson-Hiscock-Samuel analytical approximation to the RSET~\cite{Anderson:1994hg} to an expression that does not involve higher-order derivatives (thus satisfactorily implementing the desired order reduction). The situation is more involved for the RMV-RSET. Even after applying Eqs.~\eqref{eq:derrels}, we still have to solve Eqs.~\eqref{eq:phipsieqs}. However, it is likely that these differential equations are drastically simplified after the insertion of Eqs.~\eqref{eq:derrels}. Evaluating the equations for $\varphi$ and $\psi$ after application of Eqs.~\eqref{eq:derrels} is therefore a natural next step to consider, which we discuss in detail in the next section.

Before discussing the order reduction of the RMV-RSET, let us consider the interpretation of the form of the equations obtained after the application of the order-reduction procedure. These equations will take the form
\begin{equation}\label{eq:vaceqs_or}
 G_{ab}=8\pi \hat{T}_{ab}^{\rm (OR)}\,,
\end{equation}
where $\hat{T}_{ab}^{\rm (OR)}$ is the outcome of applying the order-reduction procedure to $\langle\hat{T}_{ab}\rangle$. These equations are inequivalent to Eq.~\eqref{eq:mateqs}. If seen as an approximation to the latter, Eq.~\eqref{eq:vaceqs_or} can only capture partial information about the vacuum semiclassical equations. However, Eq.~\eqref{eq:mateqs} is already an approximation to the unknown exact equations describing semiclassical backreaction, which invites interpreting these two sets of equations as two different approximations with a similar domain of validity, while potentially being self-consistent and independently well-defined. From this perspective, it is interesting to analyze different approximations to the RSET, with the aim of determining which features are universal and which ones are dependent on the particular approximation considered. In this paper, following~\cite{Arrechea:2022dvy}, we adhere to this interpretation and accept Eq.~\eqref{eq:vaceqs_or} as a possibly self-consistent set of modified gravity equations that satisfies all necessary requirements to describe semiclassical equations and is well motivated. We will compare the solutions of these vacuum semiclassical equations with the equations before application of the order-reduction procedure, and also to
other approximations to the RSET that have been previously explored (e.g.,~\cite{Fabbri:2005zn,Ho:2017joh,Berthiere:2017tms,Ho:2018jkm,Arrechea:2019jgx,Arrechea:2021ldl,Arrechea:2022dvy,Beltran-Palau:2022nec}).

\section{Order reduction of the RMV-RSET to first order \label{sec:order_red}}

\subsection{Differential equations}\label{sec:diff_eq_1st}

In this section, we discuss the application of the order-reduction procedure to the RMV-RSET. First, by using Eqs.~\eqref{eq:derrels}, the equations of motion for the scalars $\varphi$ and $\psi$ in Eqs.~(\ref{eq:phipsieqs}), are respectively reduced to
\begin{equation}\label{eq:or-phi}
   \varphi ^{(4)}+\frac{4 h}{r}\varphi^{(3)}
   +\frac{2 \left(h^2-1\right)}{r^2}\varphi{''}
   -\frac{4(h-1)}{r^3}\varphi{'}
   - \frac{6 (h-1)^2}{r^4}=O(\hbar)\,,
\end{equation}
and
\begin{equation}\label{eq:or-psi}
   \psi ^{(4)}+\frac{4 h}{r}\psi^{(3)}
   +\frac{2 \left(h^2-1\right)}{r^2}\psi{''}
   -\frac{4(h-1)}{r^3}\psi{'}
   - \frac{6 (h-1)^2}{r^4}=O(\hbar)\,,
\end{equation}
where $'$ denotes differentiation with respect to the radial coordinate, and $\varphi^{(k)}$ is the derivative of order $k\geq 3$ with respect to the radial coordinate of the field $\varphi$.

The first important observation is that $\varphi$ and $\psi$ follow the same differential equation after applying the order-reduction procedure, even though the background is not necessarily Ricci-flat. This simplifies the problem, as we just need to deal with two copies of the same differential equation. This equation includes only derivative terms of $\varphi$ and $\psi$ and does not include $\varphi$ and $\psi$ themselves. Hence, while solving Eqs.~(\ref{eq:phipsieqs}) for $\varphi$ or $\psi$ requires specifying four boundary conditions, we can solve Eqs.~(\ref{eq:or-phi}) and (\ref{eq:or-psi}) for $\varphi'$ or $\psi'$ specifying three boundary conditions only.

We will denote order-reduced tensors with the index (OR), e.g., $E_{ab}^{\rm (OR)}$. Then, the order-reduced tensors $E_{ab}^{\rm (OR)}$, $F_{ab}^{\rm (OR)}$, and $T_{ab}^{\rm (OR)}$ are given as follows:
\begin{align}\label{eq:or-set}
    T_{ab}^{\rm (OR)}=b' E_{ab}^{\rm (OR)} + b F_{ab}^{\rm (OR)}\,,
\end{align}
with components
\begin{align}
    E_{tt}^{\rm (OR)}=&
    -\frac{f}{6 r^3 h^2}\Big\{ r \left[4 (h^2-5 h+4) \varphi ''+2 r (7 h+1) \varphi ^{(3)}
    +4 r^2\varphi ^{(4)}-r^2 \varphi ''^2\right]\nonumber\\
    &-2 \varphi ' \left[- r^3 \varphi ^{(3)}+8 r^2 \varphi '' - 2 h (2 r^2 \varphi ''+9)+6h^2 +12\right]+2 r (h^2-8 h+1) \varphi'^2\Big\}\,,\label{eq:or-E-tt}\\
    E_{rr}^{\rm (OR)}=&
    \frac{1}{6 r^3h}\Big\{-2 \varphi ' \left[3 r^3 \varphi^{(3)} + h (4 r^2 \varphi ''-22)+6 h^2+24\right]+2 r (h^2-4 h+9) \varphi '^2\nonumber\\
    &+r \left[4 h^2 \varphi ''+2 h(r \varphi ^{(3)}+6 \varphi '')+3 r (2 \varphi ^{(3)}+r \varphi ''^2)\right]\Big\}\,,\label{eq:or-E-rr}\\
    E_{\theta \theta}^{\rm (OR)}=&
    \frac{1}{6 r h^2}\Big\{r \left[4 (2 h^2+h-5) \varphi ''+4 r(4 h-1) \varphi ^{(3)}+4r^2 \varphi^{(4)}-r^2 \varphi ''^2\right]\nonumber\\
    &+2 \varphi ' (6 h^2-32 h+r^3 \varphi ^{(3)}+4 r^2 \varphi''+30)-2 r (h^2-6 h+5) \varphi '^2\Big\}\,,\label{eq:or-E-thth}
\end{align}
and
\begin{align}\
    F_{tt}^{\rm (OR)}=
    &-\frac{f}{3 r^3 h^2}\Big\{ r \left[-\varphi '' (12 h+r^2 \psi ''-12)+2 (h^2-5 h+4)\psi ''\right.\nonumber\\
    &\left.+r(7 h+1) \psi ^{(3)}+2 r^2 \psi ^{(4)}\right]+\varphi ' \left[4 h(r^2 \psi ''-4 r \psi '+6)+2 h^2 (r \psi '-3)+r^3 \psi ^{(3)}\right.\nonumber\\
    &\left.-8 r^2\psi ''+2 r \psi '-18\right]+\psi ' \left[2 h (2 r^2 \varphi ''+9)-6 h^2+r^3 \varphi^{(3)}-8 r^2 \varphi ''-12\right]\Big\}\,,\label{eq:or-F-tt}\\
    F_{rr}^{\rm (OR)}=
    &\frac{1}{3 r^3 h}\Big\{\varphi ' \left[-4 h (r^2 \psi ''+2 r \psi '-6)+2 h^2 (r \psi'-3)-3 (r^3 \psi ^{(3)}-6 r \psi '+6)\right]\nonumber\\
    &-\psi ' \left[h (4 r^2\varphi ''-22)+6 h^2+3 (r^3 \varphi ^{(3)}+8)\right]\nonumber\\&+r \left[2 h^2 \psi ''+h(r \psi ^{(3)}+6 \psi '')+3 r (r \varphi '' \psi''+\psi^{(3)})\right]\Big\}\,,\label{eq:or-F-rr}\\
    F_{\theta \theta}^{\rm (OR)}=
    &\frac{1}{3 rh^2}\Big\{r \left[\varphi '' (6 h - r^2 \psi ''-6)+2 (2 h^2+h-5)\psi ''+2 r(4 h-1) \psi ^{(3)}+2r^2 \psi ^{(4)}\right]\nonumber\\
    &+\varphi ' \left[h^2(6-2 r \psi ')+12 h (r \psi '-2)+r^3 \psi ^{(3)}+4 r^2 \psi ''-10 r\psi '+18 \right]\nonumber\\
    &+\psi ' (6 h^2-32 h+r^3 \varphi ^{(3)}+4 r^2 \varphi ''+30)\Big\}\,.\label{eq:or-F-thth}
\end{align}

The traces of $E^{\rm (OR)}_{ab}$ and $F^{\rm (OR)}_{ab}$ give the order-reduced differential equations for $\varphi$ and $\psi$, i.e., Eqs.~(\ref{eq:or-phi}) and (\ref{eq:or-psi}).

The order-reduced tensors $E^{\rm (OR)}_{ab}$ and $F^{\rm (OR)}_{ab}$ are not covariantly conserved under the OR approximation. This means that the combination of the temporal and radial components of the Einstein equations is not compatible with the angular ones. From the perspective of perturbative order-reduction, this is not a problem as non-conservation appears at higher orders in the expansion. However, from the perspective of constructing a self-consistent theory of second order in derivatives, it is interesting to understand whether this issue can be solved. Conservation can indeed be restored by introducing compensatory terms, which provide a self-consistent set of differential equations~\cite{Arrechea:2022dvy} (an alternative possibility that has not been analyzed so far would be using a third auxiliary field, which is sometimes introduced in this formalism~\cite{Anderson:2007eu,Arrechea:2024ajt}, to restore conservation). To determine the form of this compensatory term, we only need to consider the single non-trivial component of the conservation equation:
\begin{equation}
\nabla_a T^{a{\ \rm (OR)}}_{\ r}=\partial_r T^{r{\ \rm (OR)}}_{\ r}+\frac{2}{r}\left(T^{r{\ \rm (OR)}}_{\ r}-T^{\theta{\ \rm (OR)}}_{\ \theta} \right)+\frac{f'}{2f}\left(T^{r{\ \rm (OR)}}_{\ r}-T^{t{\ \rm (OR)}}_{\ t} \right)\,.
\end{equation}
It follows that it is always possible to add a term $\Delta T_{\theta\theta}^{\rm (OR)}$, replacing $T_{\theta\theta}^{\rm (OR)}\rightarrow T_{\theta\theta}^{\rm (OR)}+\Delta T_{\theta\theta}^{\rm (OR)}$ and solving algebraically for  $\Delta T_{\theta\theta}^{\rm (OR)}$ to guarantee conservation:
\begin{equation}\label{eq:compdef}
\frac{2}{r}\Delta T^{\theta{\ \rm (OR)}}_{\ \theta}=\partial_r T^{r{\ \rm (OR)}}_{\ r}+\frac{2}{r}\left(T^{r{\ \rm (OR)}}_{\ r}-T^{\theta{\ \rm (OR)}}_{\ \theta} \right)+\frac{f'}{2f}\left(T^{r{\ \rm (OR)}}_{\ r}-T^{t{\ \rm (OR)}}_{\ t} \right)\,.
\end{equation}
This form of the angular component of the order-reduced stress-energy tensor is implicitly assumed when keeping only the temporal and radial components of the Einstein equations for their integration, which is the procedure followed below. Due to the introduction of compensatory terms, Eqs.~\eqref{eq:or-phi} and~\eqref{eq:or-psi} change, as determined self-consistently through Eq.~\eqref{eq:phipsieqs}, to:
\begin{align}
    & g^{ab} E^{\ \rm (OR)}_{ab} + 2 \Delta E^{\theta{\ \rm (OR)}}_{~\theta}= \frac{12 (h-1)^2}{r^4 h^2} + O(\hbar)\,,
    \\
    &g^{ab} F^{\ \rm (OR)}_{ab} + 2 \Delta F^{\theta{\ \rm (OR)}}_{~\theta} = \frac{12 (h-1)^2}{r^4 h^2} + O(\hbar)\,,
\end{align}
where the quantities of $\Delta E^{\theta{\ \rm (OR)}}_{~\theta}$ and $\Delta F^{\theta{\ \rm (OR)}}_{~\theta}$ are defined analogously to Eq. \eqref{eq:compdef}. These equations can be written as
\begin{align}\MoveEqLeft[3]\label{eq:or-phi_comp}
\begin{autobreak}
   \varphi ^{(4)}+\frac{4 h}{r}\varphi^{(3)}+\frac{2 (h^2-1)}{r^2}\varphi''  -\frac{4(h - 1)}{r^3}\varphi' - \frac{6 (h-1)^2}{r^4}
   -\frac{1}{r^4 h \left[rf'-f (h-3 r \varphi '+5)\right]}\Big\{ r^3 \varphi ^{(3)} ( r h f' (h-2 r \varphi '+1)-f (h (r h'+8 r \varphi '-7)
   -6 r h'(r \varphi '-1)+h^2 (5-8 r \varphi ')+2 h^3) )+r h f' (2 h^2 (r^2 \varphi''-3)+h (-2 \varphi ' (2 r^3 \varphi ''+r)
   +8 r^2 \varphi ''+2 r^2 \varphi '^2+12)+r^4 \varphi ''^2+\varphi' (4 r^3 \varphi ''-2 r)-6 r^2 \varphi ''+4 r^2 \varphi '^2-6)
   -f (3 r^2 h' (r^3 \varphi ''^2+6 r \varphi'^2-16 \varphi ')+h (6 r^2 (r h'+1) \varphi ''-2 r^2 (2 r h'+11) \varphi '^2
   +2 r \varphi '(h' (11 r-2 r^3 \varphi '')-2 r^2 \varphi ''+25)-4 r^4 \varphi ''^2+6)+2 h^4 (r^2 \varphi''-3)
   -2 h^3 (-6 r^2 \varphi ''+r^2 \varphi '^2+2 r \varphi ' (2 r^2 \varphi ''-5)-9)+2 h^2 (2r^4 \varphi ''^2-10 r^2 \varphi ''
   +12 r^2 \varphi '^2+r \varphi ' (6 r^2 \varphi ''-35)-9)) \Big\}=O(\hbar)\,,
\end{autobreak}
\end{align}
and
\begin{align}\MoveEqLeft[3]\label{eq:or-psi_comp}
\begin{autobreak}
   \psi ^{(4)}+\frac{4 h}{r}\psi^{(3)}+\frac{2 (h^2-1)}{r^2}\psi'' - \frac{4(h - 1)}{r^3}\psi' - \frac{6 (h-1)^2}{r^4}
   -\frac{1}{r^4 h (r f'-f (h-3 r \varphi '+5))}\Big[ r^3 \psi ^{(3)} (r h f' (h-2 r \varphi '+1)-f (h (r h'+8 r \varphi '-7)
   -6 r h'(r \varphi '-1)+h^2 (5-8 r \varphi ')+2 h^3))+6 (r (h-1) h f' (-h+r^2 \varphi''+1)+f (6 r^2 h' \varphi '
   - h (2 r (2 r h'+3) \varphi '+r^2 \varphi ''+1)-h^3(r^2 \varphi ''+4 r \varphi '+3)+h^2 (2 r^2 \varphi ''+10 r \varphi '+3)+h^4))
   +2 r^2 \psi ''(r h f' (h (4-2 r \varphi ')+h^2+r^2 \varphi ''+2 r \varphi '-3)-f (3 r^3 h' \varphi''+h (r h' (3-2 r \varphi ')
   -4 r^2 \varphi ''-2 r \varphi '+3)+2 h^2 (2 r^2 \varphi ''+3 r \varphi'-5)+h^3 (6-4 r \varphi ')+h^4))
   +r \psi ' (f (6 r h' (r^3 \varphi ^{(3)}-6 r \varphi '+8)+h (2 r h' (2 r^2 \varphi ''+4 r \varphi '-11)-3 r^4 \varphi ^{(4)}
   -8 r^3 \varphi ^{(3)}+10 r^2\varphi ''+32 r \varphi '-32)+2 h^3 (r^2 \varphi ''+2 r \varphi '-1)
   -2 h^2 (2 r^3 \varphi ^{(3)}+6 r^2 \varphi''+18 r \varphi '-17))-2 r h f' (h (2 r^2 \varphi ''-2 r \varphi '+1)
   +r^3 \varphi ^{(3)} - 2 r^2\varphi ''-4 r \varphi '+1)) \Big]=O(\hbar)\,.
\end{autobreak}
\end{align}
Note that the terms on the second and following lines of the left-hand side of Eqs.~\eqref{eq:or-phi_comp} and~\eqref{eq:or-psi_comp} vanish on the Schwarzschild background.

This procedure is not unique, since there are different ways of restoring conservation at the lowest order. This issue can be understood in two different ways:
\begin{itemize}
\item From the perspective of perturbative order-reduction, the failure of conservation signals the breakdown of the derivative expansion. That is, the order-reduced equations form part of an approximation scheme that can break down in certain regimes. In fact, the terms in the square brackets of Eqs.~\eqref{eq:or-phi_comp} and \eqref{eq:or-psi_comp} vanish for the Schwarzschild background. Therefore, they are $\mathcal{O}(\hbar)$ in the regime of validity of the derivative expansion. From this perspective, non-uniqueness is not an issue as it arises in a regime that is beyond the regime of validity of the original equations.
\item Alternatively, by first applying order reduction and afterwards introducing compensatory terms, it is possible to define second-order theories that are mathematically self-consistent. These theories will likely present sizeable deviations from the original equations in the same regimes in which the derivative expansion breaks down. From this perspective, non-uniqueness is not an issue, as possible differences between the corresponding solutions in different theories are limited to these regimes. While well-defined on mathematical grounds, this is an extrapolation of the original approximation that can significantly change the features of the resulting solutions.
\end{itemize}

We see that, in general and regardless of the introduction of compensatory terms, we expect the order-reduced RMV-RSET in Eq.~\eqref{eq:or-set} to present deviations with respect to the original RMV-RSET as defined in Eq.~\eqref{eq:rmv_def}. As a consequence, the order-reduced RMV-RSET is not guaranteed to encode the correct physics, although it might be able to provide the right qualitative insights. Hence, given the difficulties in solving the problem without the order-reduction procedure, it is reasonable to give the order-reduced RMV-RSET the benefit of the doubt and analyze the properties of the corresponding solutions, keeping this issue in mind.\footnote{A comparison with the way the solutions to the Einstein field equations are treated is illustrative. For instance, black hole solutions are pushed to their limits and studied arbitrarily close to singularities or beyond Cauchy horizons through the use of analytical extensions. That these solutions are expected to break down in these regimes does not prevent the existence of a rich body of literature analyzing these extreme situations.}

As a first non-trivial estimation of the ability of the order-reduced RMV-SET to encode the correct physics, we can evaluate Eq.~\eqref{eq:or-set} on the Schwarzschild spacetime, and compare it against Eq.~\eqref{eq:rmv_def} [which we can evaluate using the analytical expressions in Eqs.~\eqref{eq:sol-phi-Sch} and \eqref{eq:sol-phi-Sch_complete}]. For the Schwarzschild background, Eqs.~\eqref{eq:derrels} are exact without the need to introduce $\mathcal{O}(\hbar)$ terms, which in turn implies that these three approximations (namely full RMV-RSET, and order-reduced RMV-RSET  with and without compensatory terms) are degenerate and thus give exactly the same results when evaluated on the Schwarzschild background.\footnote{For Ricci-flat backgrounds, the only curvature tensor appearing in the equations of motion for the auxiliary fields, as well as the tensor components in Eqs.~\eqref{eq:Edef} and~\eqref{eq:Fdef}, is the Weyl tensor. For the Schwarzschild background, Eqs.~\eqref{eq:derrels} are exact, meaning that all components of the Weyl tensor can be written as a polynomial in $h(r)$. As a result, the RMV-RSET does not contain derivatives of the metric functions, which means that the order-reduction procedure yields the same RSET. The latter being conserved, compensatory terms defined in Eq.~\eqref{eq:compdef} vanish identically for the Schwarzschild background. This shows the degeneracy of the three approximations when evaluated on Schwarzschild backgrounds.} Note that these approximations will generally lead to different spacetimes when backreaction is included, as discussed below. However, this non-trivial result shows that the order-reduction procedure does not erase the correct perturbative (without backreaction) physics near the gravitational radius (in particular, the divergent behavior of anomaly-induced terms~\cite{Mottola:2006ew,Mottola:2025fhl} due to the blueshifting of frequencies near the horizon), which is a reasonable necessary condition for the order-reduced RMV-RSET to encode the correct non-perturbative physics as well.

\subsection{Numerical results}

In Sec.~\ref{sec:diff_eq_1st} above, we introduced a system of five differential equations. Three of these equations come from the Einstein field equations, while the remaining two equations come from the differential equations for $\varphi$ and $\psi$. Due to the Bianchi identity satisfied by curvature tensors, not all the equations obtained from the Einstein field equations are independent, and we are free to ignore the angular equations as these contain redundant information. The boundary conditions are specified by assuming that the asymptotic solutions take the form of those in the Schwarzschild spacetime, together with the higher-order corrections described in App.~\ref{app:boundary_conditions}.

In the following, we discuss the numerical results obtained when integrating the differential equations described in Sec.~\ref{sec:diff_eq_1st} with the asymptotic boundary conditions in Eq.~\eqref{eq:boundary_cond}. In the region in which the spacetime metric is well described by the Schwarzschild metric ($r \gg M$), the auxiliary fields $\varphi(r)$ and $\psi(r)$ follow the same differential equation, and the solutions are given by the form of Eq.~\eqref{eq:sol-phi-Sch}. We use the integration constants for the Boulware state, specified in Eqs.~(\ref{eq:cd-infty}-\ref{eq:ch_dh}), and start integrating from an outer boundary radius $r=r_0$.

Under the above setup, we solve the system of differential equations and obtain numerical values for the functions $f(r)$, $h(r)$, $\varphi(r)$, and $\psi(r)$ in terms of the radial coordinate $r$, with and without compensatory terms.\footnote{To obtain higher-order derivatives of the metric functions and the auxiliary fields without numerically differentiating the solution, we differentiate the semiclassical Einstein equations with respect to $r$ before numerical integration and decompose the resulting higher-order equations into a system of first-order differential equations.}

\begin{figure}[htbp]
\includegraphics[width=0.55\linewidth]{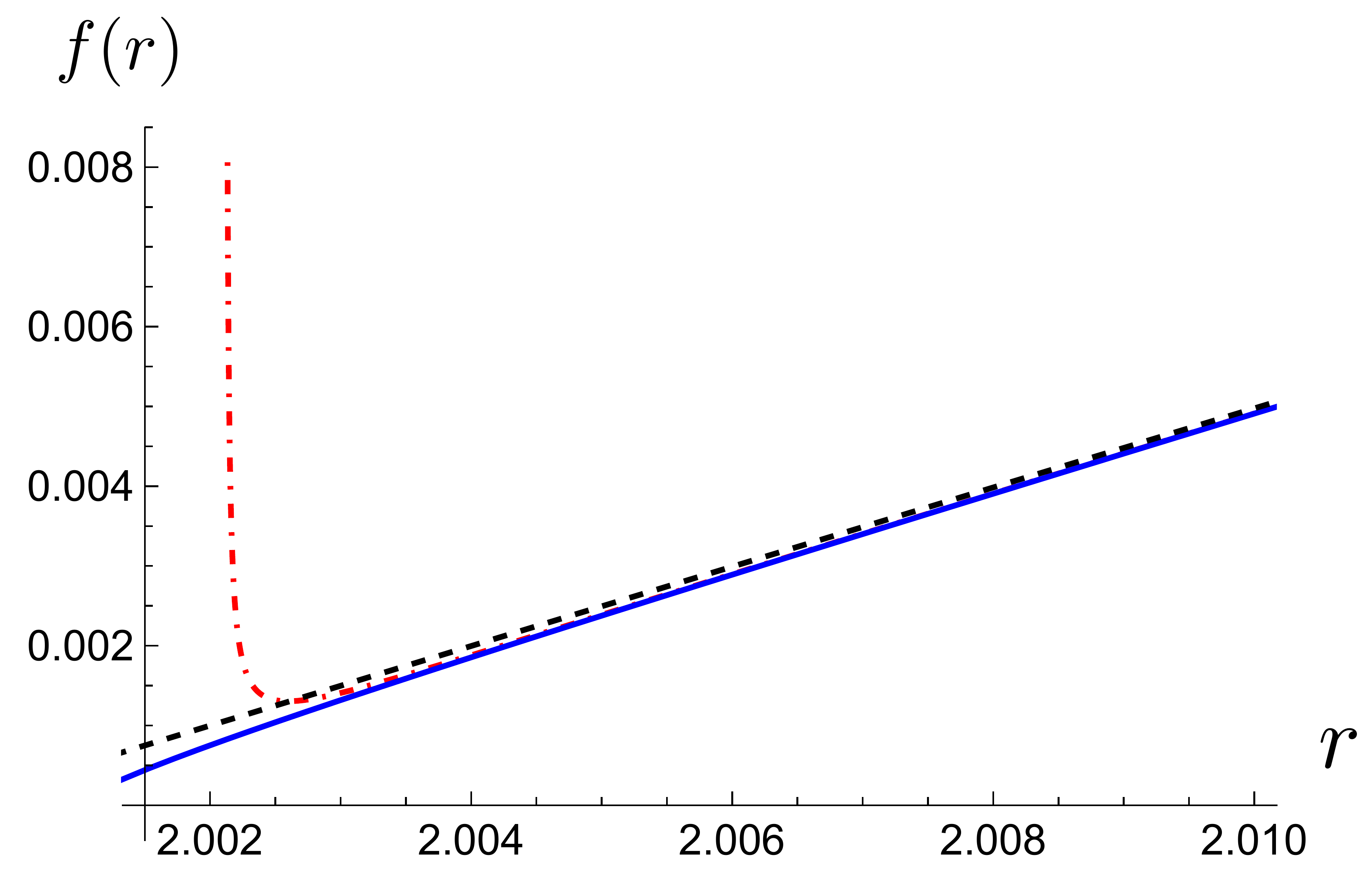}
\caption{Plots of the metric function $f(r)$ for the semiclassical solutions with (solid blue line) and without (dashed-dotted red line) compensatory terms, as well as the classical Schwarzschild solution (black dashed line).  Using numerical values $M=1$, $\hbar=10^{-2}$, and $r_0=10^2$, in the case without compensatory terms our numerical integrations show a tendency of $f(r)$ to diverge at $r \simeq 2.00211$, while introducing compensatory terms makes $f(r)$ tend towards a vanishing value at $r \simeq 2.00112$.\label{fig:sol_f}}
\end{figure}

\begin{figure}[htbp]
\includegraphics[width=0.55\linewidth]{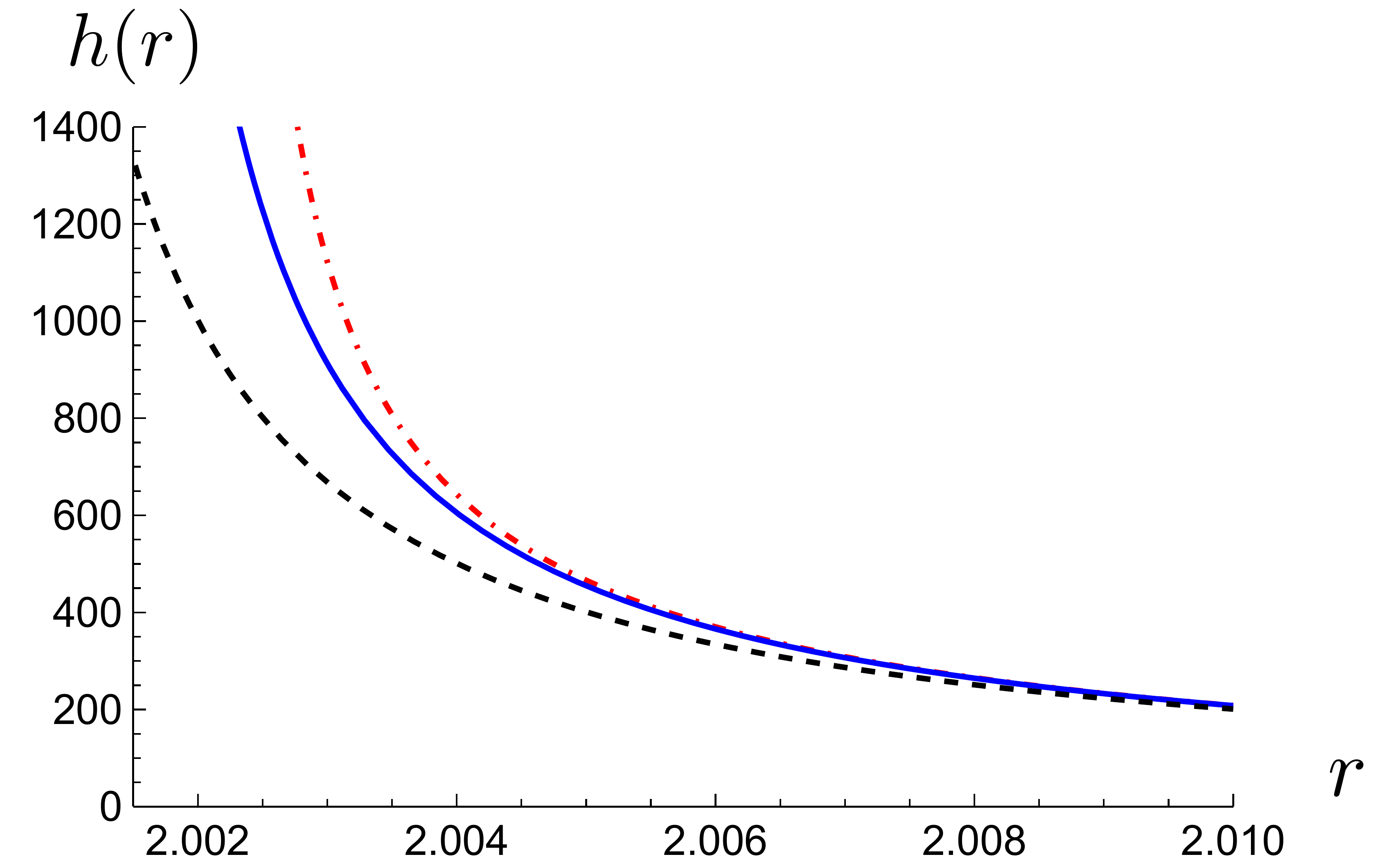}
\caption{\label{fig:sol_h}
Plots of the metric function $h(r)$ for the semiclassical solutions with (solid blue line) and without (dashed-dotted red line) compensatory terms, as well as the classical Schwarzschild solution (dashed black line). The qualitative behavior of $h(r)$ is the same as in the classical case, but with $1/h(r)$ vanishing for a slightly larger value of $r \simeq 2.00211$ without compensatory terms and $r \simeq 2.00112$ with compensatory terms (for $M=1$, $\hbar=10^{-2}$, and $r_0=10^2$).
}
\end{figure}

\begin{figure}[htbp]
\includegraphics[width=0.55\linewidth]{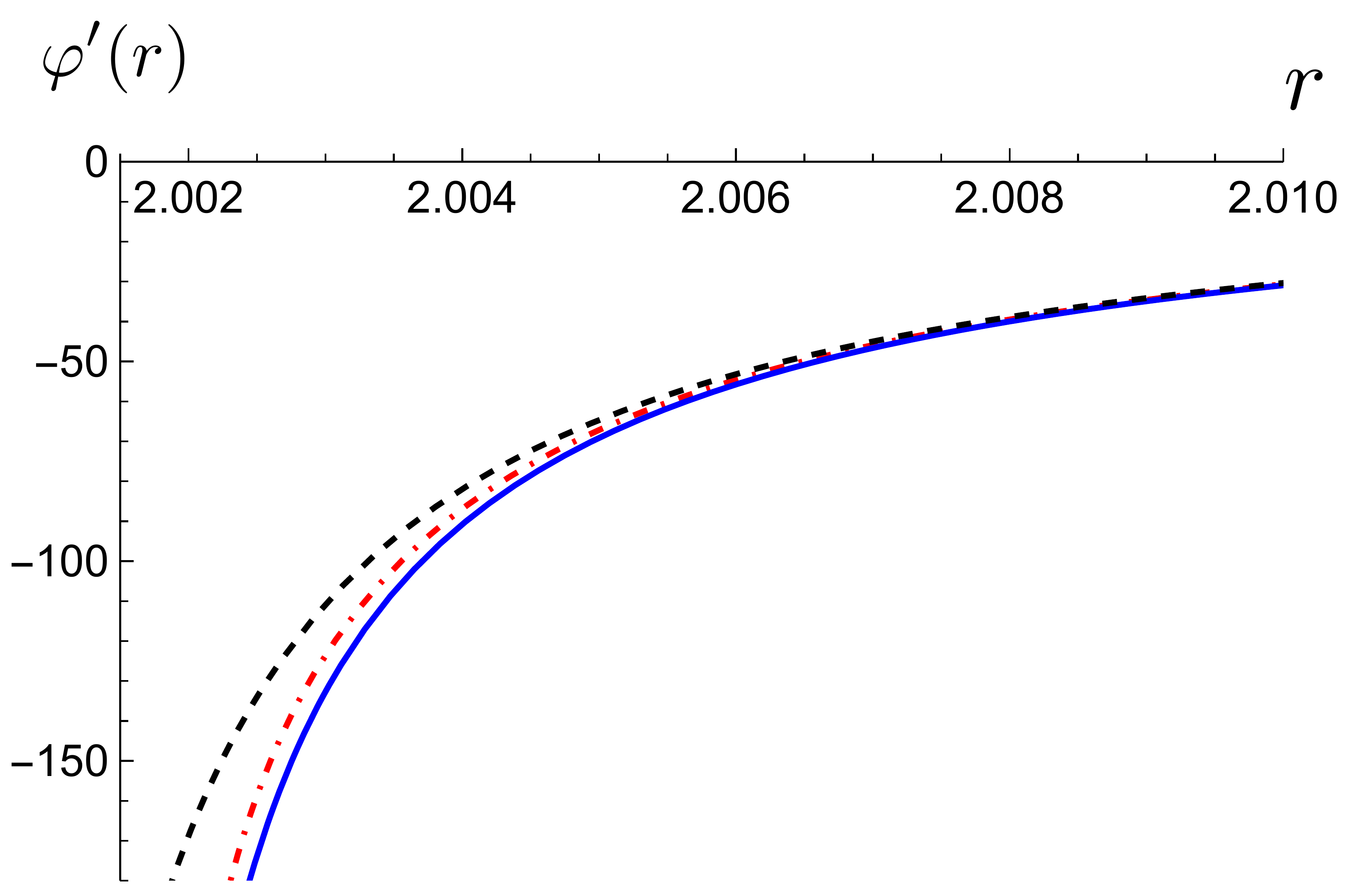}
\includegraphics[width=0.55\linewidth]{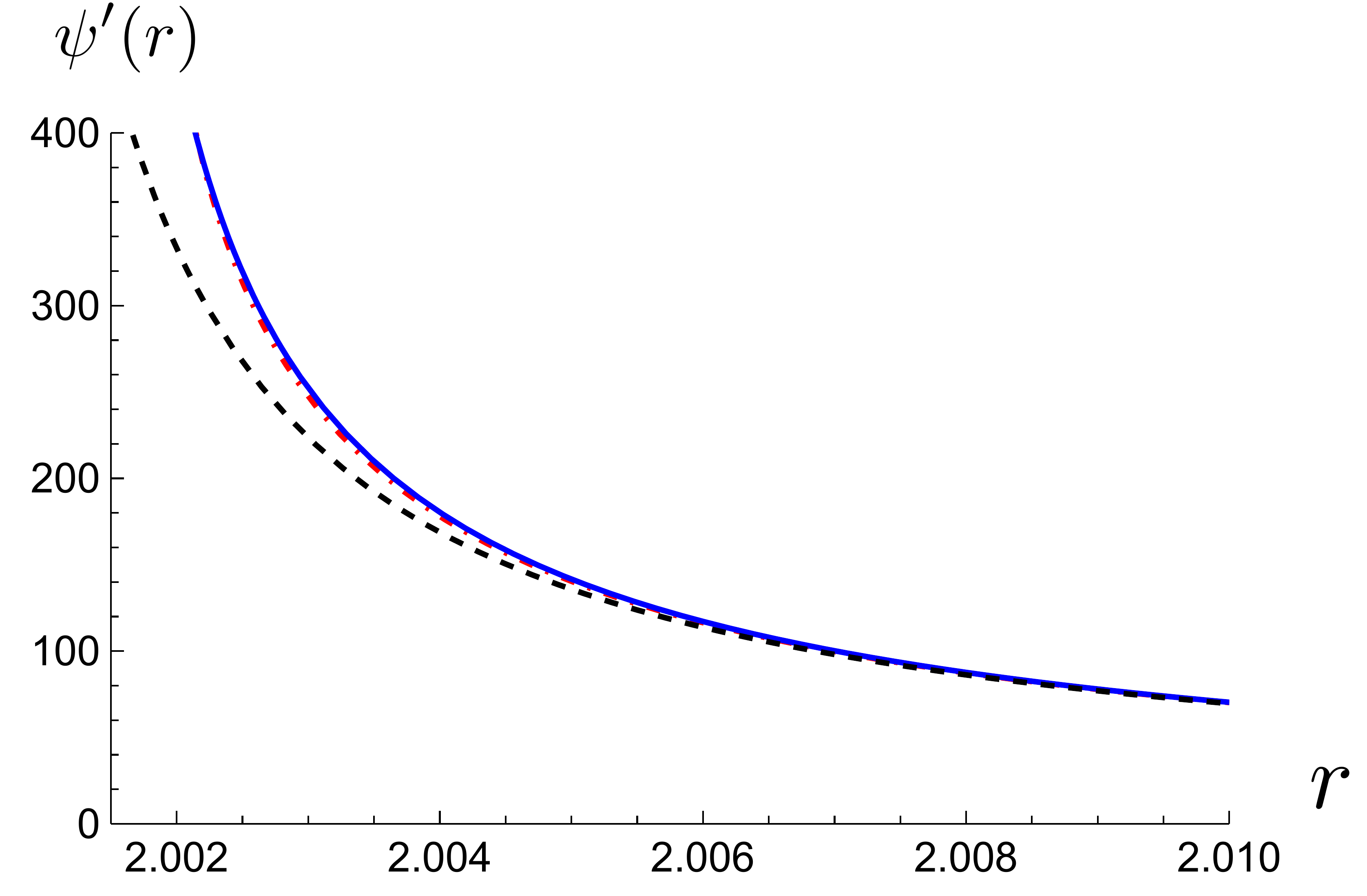}
\caption{Plots of the first $r$-derivatives of the auxiliary fields $\varphi(r)$ and $\psi(r)$ for the semiclassical solutions with (solid blue line) and without (dashed-dotted red line) compensatory terms (for $M=1$, $\hbar=10^{-2}$, and $r_0=10^2$), as well as the solution without backreaction on the Schwarzschild spacetime (dashed black line), given by Eq.~\eqref{eq:sol-phi-Sch}. The numerical values obtained by solving the equations with backreaction differ from the analytic expression in Eq.~\eqref{eq:sol-phi-Sch} close to the gravitational radius, which indicates that backreaction effects become important there.\label{fig:sol_dphi}}
\end{figure}

These results will be compared with the classical Schwarzschild solution, as a reference. The metric functions are given in Figs.~\ref{fig:sol_f} and~\ref{fig:sol_h}, while the derivatives of the fields $\varphi(r)$ and $\psi(r)$ are presented in Figs.~\ref{fig:sol_dphi}.

The behavior we obtained is the following. Our numerical simulations become singular at specific values of the radius $r$, which are related to the behavior of the metric functions $f(r)$ and $h(r)$. This behavior is different depending on whether or not compensatory terms are included (see Fig.~\ref{fig:sol_f}). In the absence of compensatory terms, the metric function $f(r)$ grows steadily until the minimum radius is reached by our simulations, $r=r_{\rm cr,1}\simeq 2M\left[1+K_1\sqrt{\hbar}/2M\right]$, with $K_1\simeq 2.16 \times 10^{-2}$. When introducing compensatory terms, $f(r)$ decreases steadily towards a vanishing value as the radius approaches $r=r_{\rm cr,2}\simeq 2M\left[1+K_2\sqrt{\hbar}/2M\right]$, with $K_2\simeq 1.12 \times 10^{-2}$. We cannot determine whether $f(r)$ diverges or vanishes, as both situations would imply singularities around which numerical simulations break down, but we can determine tendencies within the integration domain. Whether these singularities are curvature singularities or coordinate singularities will be discussed in the paragraph below. On the other hand, the function $h(r)$ remains divergent as in the classical case (see Fig.~\ref{fig:sol_h}), although the value of the radius for which $1/h(r)$ vanishes is shifted to $r=r_{\rm cr,1}$ without compensatory terms and $r=r_{\rm cr,2}$ with compensatory terms. For completeness, we also compare the numerical solutions for $\varphi'(r)$ and $\psi'(r)$ with the analytic solution in Schwarzschild spacetime, Eq.~\eqref{eq:sol-phi-Sch} (see Fig.~\ref{fig:sol_dphi}), which provides an alternative representation of the regions in which semiclassical backreaction effects become important.

\begin{figure}[htbp]
\includegraphics[width=0.55\linewidth]{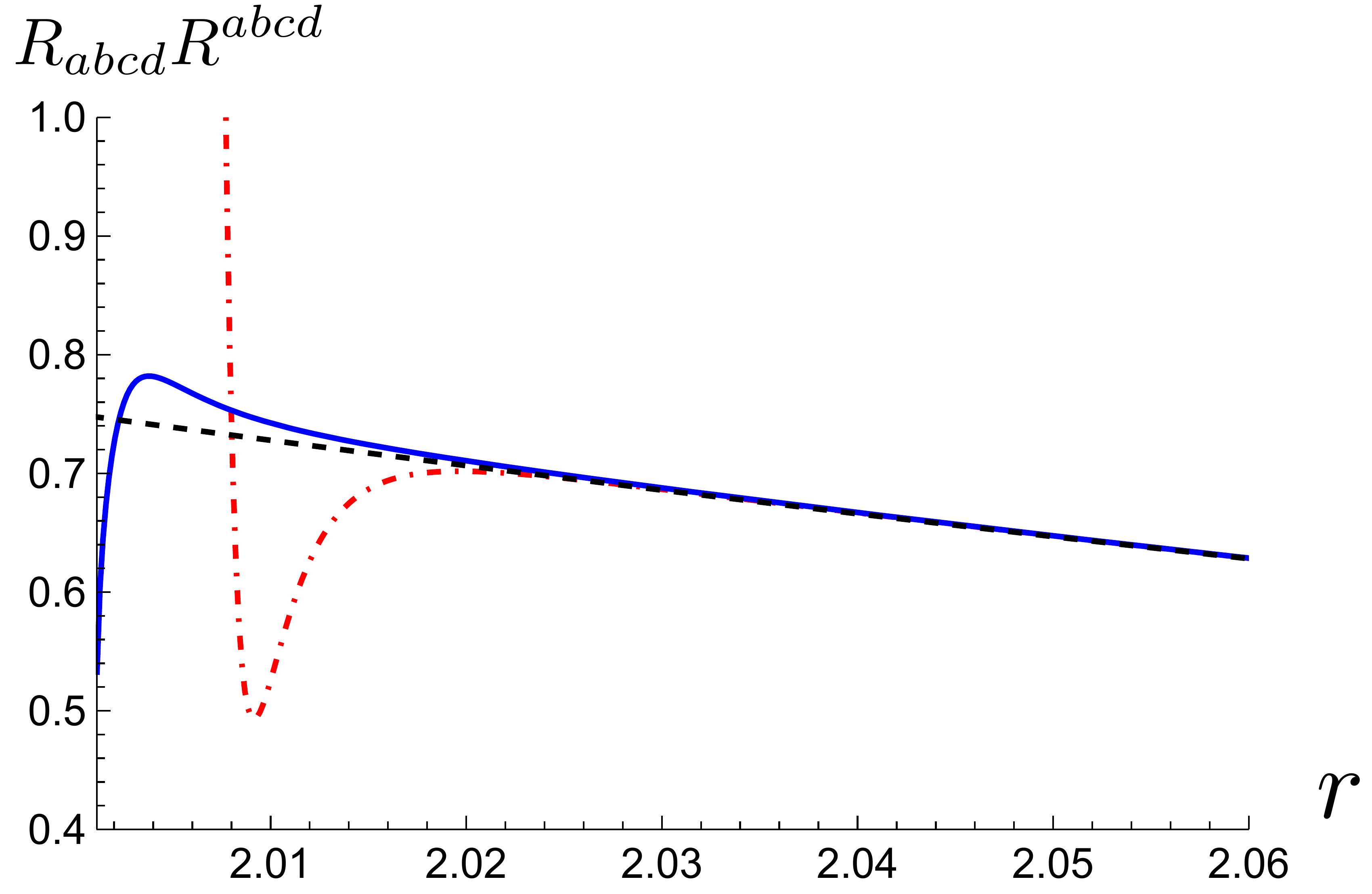}
\caption{Kretschmann scalar evaluated for the Schwarzschild metric (dashed black line) and the order-reduced semiclassical Einstein equation with (solid blue line) and without (dashed-dotted red line) compensatory terms. Taking into account that the Kretschmann scalar is positive definite, the observed behavior is compatible with a divergence as the radius approaches $r \simeq 2.00211$ for the solutions without compensatory terms, while this quantity remains finite as the radius approaches $r \simeq 2.00112$ for the solution with compensatory terms (for $M=1$, $\hbar=10^{-2}$, and $r_0=10^2$).
\label{fig:curvature_inv}}
\end{figure}

The behavior of the metric functions may indicate a spacetime singularity, or it may be a coordinate singularity. For a static and spherically symmetric spacetime, finiteness of the Kretschmann scalar suffices to guarantee regularity~\cite{Bronnikov:2012wsj} (this result actually holds for static spacetimes without any further symmetry requirements~\cite{Lobo:2020ffi}). This follows from the observation that the Kretschmann scalar can be written as a sum of squares which, for the line element in Eq.~\eqref{eq:sssansatz}, takes the form:
\begin{equation}
K=\frac{4}{r^4}\left\{r^2\frac{\left(h'\right)^2}{2h^2}+r^2\frac{\left(f'\right)^2}{2f^2h^2}+\left(1+\frac{h'}{h^2}\right)^2+r^4\left[\frac{f''}{2fh}-\frac{f'h'}{4fh^2}-\frac{\left(f'\right)^2}{4f^2h}\right]^2\right\}\,.
\end{equation}
The behavior of the Kretschmann scalar for the numerical integrations discussed above is shown in Fig.~\ref{fig:curvature_inv}. We can see that the solution without compensatory terms has a divergent Kretschmann scalar, while this quantity remains finite for the solution with compensatory terms. The existence of
singularities in some cases may be connected to the order-reduction procedure, as discussed in the next section. For completeness, we analyze semiclassical backreaction under an equivalent metric ansatz in App. \ref{app:alt_ansatz} to check the validity of the numerical analysis.

The next geometric aspect we will explore is whether or not these singularities are naked. For this purpose, let us introduce horizon-penetrating coordinates allowing for a clean comparison with the Schwarzschild spacetime. In particular, let us consider the straightforward generalization of the ingoing Eddington-Finkelstein coordinates, starting with the definition of the coordinate $r^*$ by the relation
\begin{equation}
\text{d}r^*=\mbox{sgn}[f(r)]\sqrt{\frac{h(r)}{f(r)}}\text{d}r\,,
\end{equation}
which is well-defined everywhere for our spacetimes as both $f(r)$ and $h(r)$ remain positive, as well as the null coordinate
\begin{equation}
v=t+r^*,
\end{equation}
so that the line element in Eq.~\eqref{eq:sssansatz} becomes, in the $(v,r)$ coordinates,
\begin{equation}\label{eq:sssansatzv}
\text{d}s^2=-f(r)\text{d}v^2+2\mbox{sgn}[f(r)]\sqrt{f(r)h(r)}\text{d}v\text{d}r+r^2\text{d}\Omega^2\,.
\end{equation}
This line element has the following radial null vector fields
\begin{equation}
\bm{n}=-\partial_r,\qquad \bm{l}=\partial_v+\frac{\mbox{sgn}[f(r)]}{2}\sqrt{\frac{f(r)}{h(r)}}\partial_r\,,
\end{equation}
satisfying the relations
\begin{equation}
\bm{n}^2=\bm{l}^2=0,\qquad \bm{n}\cdot\bm{l}=-1\,.
\end{equation}
The standard calculation of the corresponding expansions yields the results
\begin{equation}
\theta_{\bm{n}}=-\frac{2}{r},\qquad \theta_{\bm{l}}=\frac{\mbox{sgn}[f(r)]}{r}\sqrt{\frac{f(r)}{h(r)}}\,.
\end{equation}
From these expressions, we can read that the expansion $\theta_{\bm{n}}$ is negative throughout the spacetime, actually having the same value as in flat spacetime, while the value of the expansion $\theta_{\bm {l}}$ is a function of the metric components. For the geometries considered in this paper, $\theta_{\bm {l}}$ remains positive, and thus these geometries do not contain trapping horizons (since the spacetimes considered are static, this also implies the lack of event horizons). This can be compared with the result for the Schwarzschild spacetime in which $\left.\theta_{\bm{l}}\right|_{r=2M}=0$.

Situations in which $h(r)$ diverges without $f(r)$ simultaneously vanishing correspond generically to wormhole metrics. In the numerical integrations detailed above, $f(r)$ either diverges (without compensatory terms) or tends to vanish (with compensatory terms) for the same value of the radius, at least within the numerical resolution with which the integrations have been performed. For the former case, we have also observed a divergent tendency of the Kretschmann scalar at $r=r_{\rm cr,1}$, compatible with the existence of a curvature singularity at the would-be throat. For the latter, the Kretschmann scalar tends to a finite value. Interpreting the latter case is more subtle, as regardless of whether the metric would have a horizon or a wormhole throat at $r=r_{\rm cr,2}$, the coordinates used would cease to work there. Increasing the numerical precision did not solve the problem.

However, we can particularize more general results to our current setup in order to resolve numerical ambiguities and to constrain the resulting geometries~\cite{Carballo-Rubio:2025fhq}. In general, the hypersurface $r=r_{\rm cr,2}$ in the integrations with compensatory terms can either be a black hole horizon or a wormhole throat, depending on the behavior of the metric functions and the energy density of the RMV-RSET. For an adequate RSET approximation, we expect the energy density to be unbounded on Killing horizons~\cite{Carballo-Rubio:2025fhq}.
We have confirmed this expectation by performing some auxiliary numerical evaluation of the order-reduced RSET with compensatory terms, as detailed in App.~\ref{app:rhodiv}. This unbounded nature would imply a divergent Kretschmann scalar at $r=r_{\rm cr,2}$, which is not compatible with our numerical results. This strongly suggests that $r=r_{\rm cr,2}$ is a wormhole throat in the case with compensatory terms.
The sign of the first derivative of $g^{rr}$ is also compatible with a wormhole throat, as we have checked numerically that it is positive at the endpoint of numerical integration $r=r_{\rm cr,2}$~\cite{Visser:1997yn,Carballo-Rubio:2025fhq}.

\subsection{Comparison with the full RMV-RSET and previous literature}

In previous sections, we have discussed the approximation to the RSET obtained by reducing the order of the RMV-RSET to first order. The RMV-RSET and its order reduction to first order are different approximations and, for completeness, we compare them here. We also compare our results for the backreaction problem using the order-reduced RMV-SET with previous work in the literature. This comparison is especially important due to the limitations of the order-reduction procedure and the non-uniqueness in the introduction of compensatory terms discussed previously.

\begin{figure}[htbp]
\includegraphics[width=0.55\linewidth]{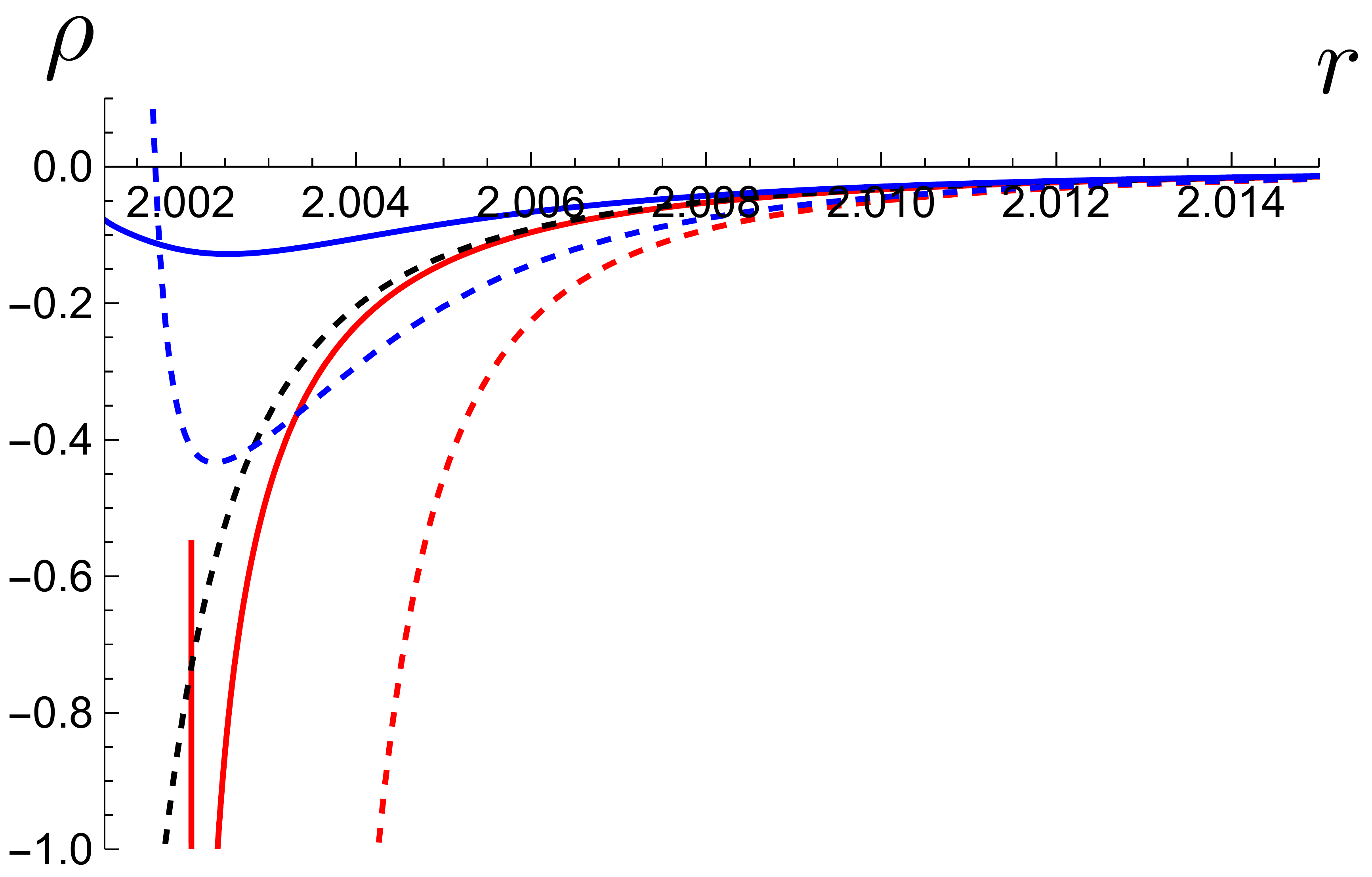}
\caption{Plot of the energy density as a function of the radius. The dashed black line indicates any of the approximations discussed in the paper (full RMV-RSET and order-reduced RMV-RSET with and without compensatory terms) evaluated on the Schwarzschild background. The solid lines indicate the order-reduced RMV-RSET in the self-consistent solutions obtained by solving the backreaction problem to first order without compensatory (red) and with compensatory (blue) terms. The remaining dashed lines indicate the full RMV-RSET evaluated on the geometries obtained when the backreaction of first order without compensatory (red) and first order with compensatory (blue) approximations are included. We see that the red (blue) solid and dashed lines have a similar behavior, and that the introduction of compensatory terms induces qualitative deviations from both the order-reduced approximations without compensatory terms and the results for a fixed Schwarzschild background.
\label{fig:rho}}
\end{figure}

\begin{figure}[htbp]
\includegraphics[width=0.55\linewidth]{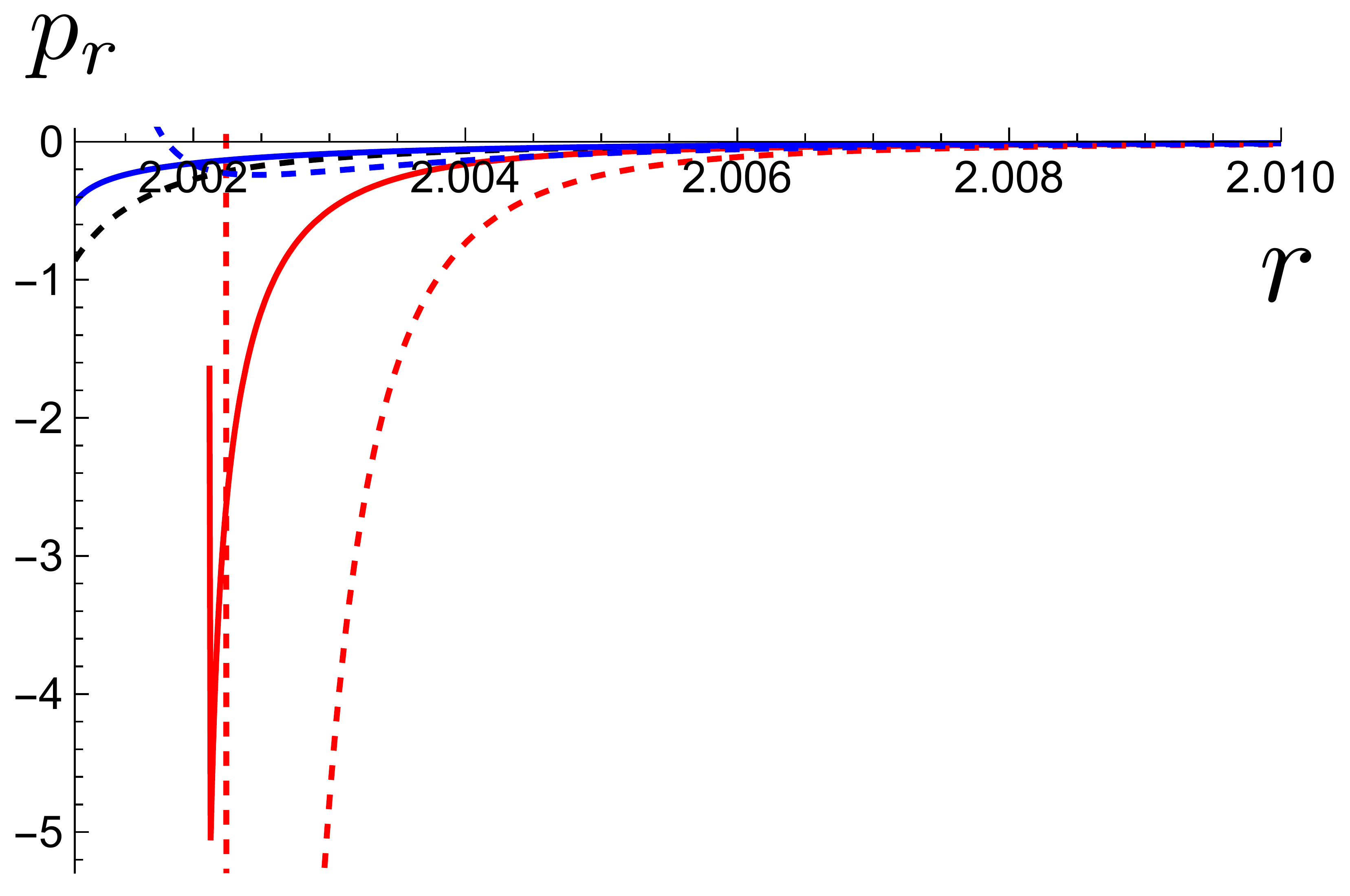}
\caption{Plot of the radial pressure as a function of the radius. Different lines have the same meaning as in Fig.~\ref{fig:rho}. For this quantity, the order-reduced stress-energy tensor with compensatory terms displays a similar behavior as the full RMV-RSET evaluated on the fixed Schwarzschild background, while the approximation without compensatory terms displays a more distinct behavior.
\label{fig:pr}}
\end{figure}

\begin{figure}[htbp]
\includegraphics[width=0.55\linewidth]{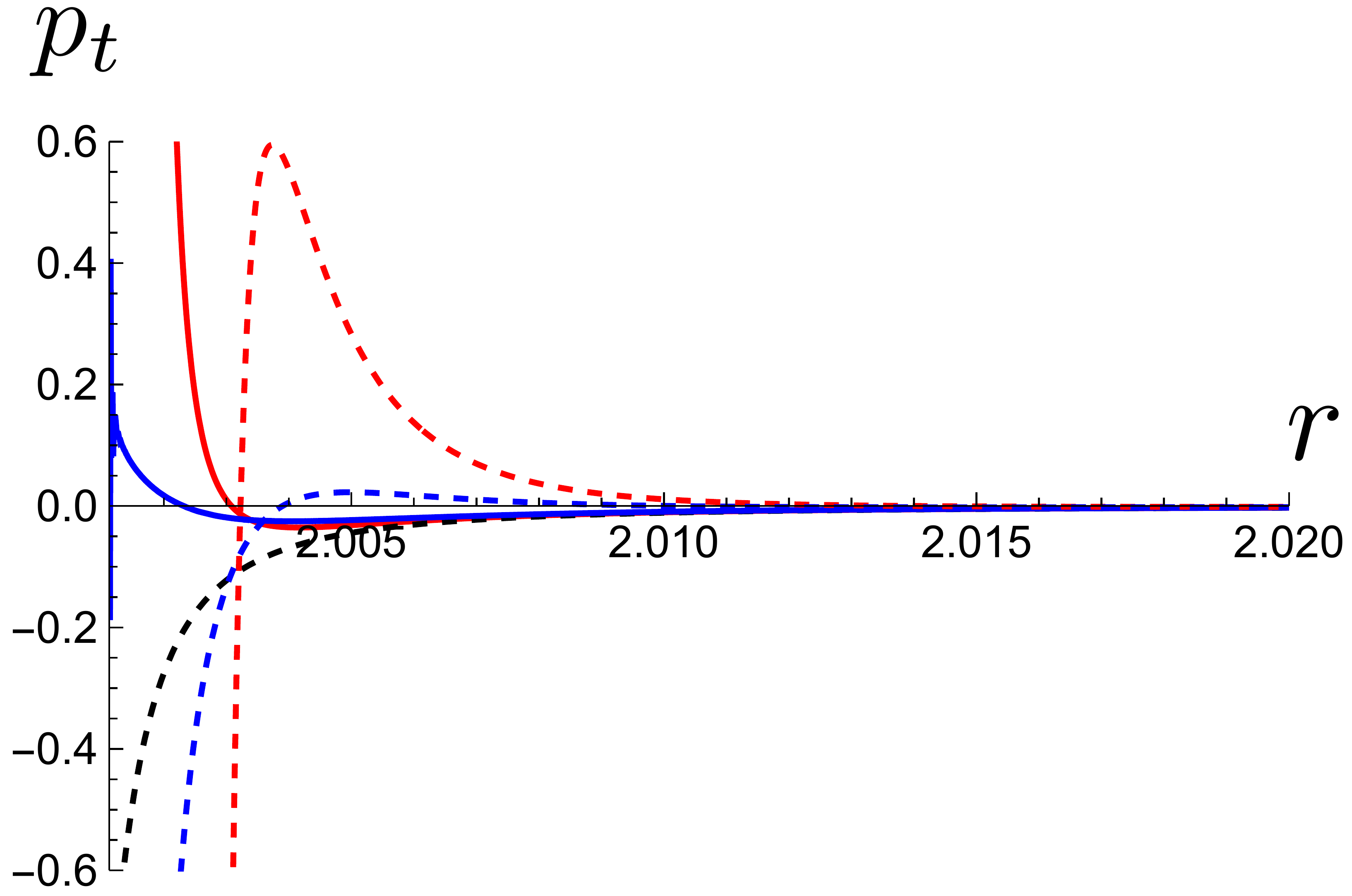}
\caption{Plot of the tangential pressure as a function of the radius. Different lines have the same meaning as in Fig.~\ref{fig:rho}. The tangential pressure is the quantity with the most clear differences for different approximations.
\label{fig:pt}}
\end{figure}

The benefit of the reduced-order RMV-RSET is that it makes the backreaction problem tractable, allowing us to obtain deformations of the Schwarzschild solution. We can plot the effective energy density and pressures associated with these deformations. On the other hand, while we have not solved the backreaction problem for the full RMV-RSET, we can evaluate this stress-energy tensor on the deformed Schwarzschild solutions obtained. This procedure results in Fig.~\ref{fig:rho} for the energy density, Fig.~\ref{fig:pr} for the radial pressure, and Fig.~\ref{fig:pt} for the tangential pressure. These quantities are defined as $\rho = -T^{t}_{~t}$, $p_{r} = T^{r}_{~r}$, and $p_{t} = T^{\theta}_{~\theta}$, respectively.


\begin{center}
\begin{table}
\begin{tabular}{|c|c|c|}
\hline
  RSET & $p_r/\rho$ & $p_t/p_r$ \\
\hline
  RMV-RSET (no backreaction)   & 1/3 & 1\\
\hline
  OR-RSET (with compensatory terms) & $5.679$ & $-0.412$ \\
\hline
  OR-RSET (without compensatory terms) & $1.349 \times 10^{4}$ & $-0.500$ \\
\hline
\end{tabular}
    \caption{Comparison of the limiting values for the ratios of the RSET components evaluated near the points in which the numerical simulations break down. Backreaction changes the values of these ratios compared to those obtained when evaluating the RMV-RSET without backreaction.}
    \label{tab:table}
\end{table}
\end{center}

The procedure of order reduction has been applied to the same problem analyzed here, but with a different form of the RSET, the so-called Anderson-Hiscock-Samuel approximation (AHS-RSET in the following), which contains an extra parameter controlling a non-minimal coupling to the spacetime curvature~\cite{Arrechea:2023oax}. The solutions obtained using the AHS-RSET belong to two classes, depending on the sign of the energy density. Whenever the energy density is negative definite, as for the RMV-RSET, the behavior of the metric functions $f(r)$ and $h(r)$ is qualitatively the same as in our integration without compensatory terms. The resulting geometries are naked singularities that can be interpreted as arising from the truncation of the full RSET due to the order-reduction procedure. From this perspective, the full RSET would give rise to solutions with regular wormhole throats that are slightly larger than the Schwarzschild radius of the asymptotic ADM mass (which has been found in several approaches~\cite{Fabbri:2005zn,Ho:2017joh,Berthiere:2017tms,Ho:2018jkm,Arrechea:2019jgx,Arrechea:2021ldl,Arrechea:2022dvy,Beltran-Palau:2022nec}), which become singular when applying the order-reduction procedure (note that these solutions are generically singular, but the singularity is not placed on the wormhole throat, but rather on the other side of it). This indicates that the procedure of order reduction can modify both regular and singular features of the solutions obtained using the full RMV-RSET. On the other hand, our integrations with compensatory terms display a different behavior that is not observed for any of the numerical solutions in~\cite{Arrechea:2023oax}, which indicates that the introduction of compensatory terms can induce important differences between different approximations.

Other works have followed a different route. The trace anomaly does not impose enough constraints to make the semiclassical system of differential equations solvable without providing further information. A possibility is then assuming the heuristic constraint $\langle p_r\rangle =\langle p_t\rangle$~\cite{Beltran-Palau:2022nec}. This heuristic constraint is based on this condition being satisfied near the gravitational radius in the absence of backreaction (see Table~\ref{tab:table}), though in~\cite{Beltran-Palau:2022nec} it is extended to all values of the radial coordinate. On the other hand, our approach does not require specifying extra conditions, as the effective action contains all the components of the RSET. We can therefore check whether the condition $\langle p_r\rangle =\langle p_t\rangle$ holds true in the presence of backreaction. In our approximation, this condition is not satisfied whenever backreaction is considered, as it can be seen in Table~\ref{tab:table}. In fact, it is reasonable to expect that, in general, this condition would break down in the presence of backreaction unless it was exact for all the values of the radial coordinate (which is not the case). Our results show that such a constraint may not be a reasonable assumption to impose when analyzing semiclassical backreaction, as the fact that it is satisfied in the absence of backreaction does not guarantee that it will hold true in the presence of backreaction.

\section{Conclusions}
\label{sec:conclusions}

The study of the backreaction of quantum fields on spacetimes of physical interest, such as the Schwarzschild solution, is an important problem that has been studied extensively under different approximation schemes. The main reason behind the proliferation of diverse approximation schemes is the complexity of the calculation of the renormalized stress-energy tensor of quantum fields on curved backgrounds. Given that the validity of different approximations, as well as their interplay with each other, is difficult to assess a priori, it is important to study specific examples and compare the results obtained.

Here, we have carried out the study of the backreaction of the Riegert--Mottola--Vaulin Renormalized Stress-Energy Tensor (RMV-RSET) after applying an order-reduction procedure. The order-reduction procedure implemented here is different from the ones considered previously due to the fact that the RMV-RSET is written in terms of auxiliary fields satisfying fourth-order equations of motion. Nevertheless, we have been able to apply such a procedure and perform numerical integrations to obtain the corresponding modifications of the Schwarzschild solution.

We have compared our results with the existing literature, pointing to the existence of aspects shared across approximations that can lead to the emergence of universal behaviors. The results obtained in the approximation without compensatory terms suggest that the order-reduction procedure can generate singularities in regimes in which the full RSET would generate non-perturbative corrections. For the solutions analyzed here, these non-perturbative corrections are expected to modify the geometry so as to produce a wormhole throat. On the other hand, we have seen how these singularities can be removed when introducing compensatory terms, which indicates that the resulting approximation to the full RMV-RSET can have significant differences with respect to the approximation without compensatory terms (which is arguably not surprising due to the latter approximation being an extrapolation which is expected to become significantly different in the near-horizon region). This provides additional motivation to study the solutions that result from the backreaction of the full RMV-RSET.

In summary, our work illustrates the importance of performing studies of semiclassical backreaction under different approximations with the aim of critically comparing results and identifying robust features across different approximations. Regardless of whether these approximations are accurate or complete, identifying such robust features could be important to determine potentially observable effects. From this perspective, a natural extension of this work would be applying the same techniques to the analysis of the backreaction of more complete expressions of the renormalized stress-energy, including in particular Weyl-invariant contributions to the latter.

\begin{acknowledgments}
The authors are thankful to Julio Arrechea for insightful comments on a previous version of the draft. RCR acknowledges financial support provided by the Spanish Government through the Ram\'on y Cajal program (contract RYC2023-045894-I) and the Grant No.~PID2023-149018NB-C43 funded~by MCIN/AEI/10.13039/501100011033, and by the Junta de Andaluc\'{\i}a
through the project FQM219 and from the Severo Ochoa grant
CEX2021-001131-S funded by MCIN/AEI/ 10.13039/501100011033, as well as the hospitality of the Center of Gravity, a Center of Excellence funded by the Danish National Research Foundation under grant No.~184. The work of FDF is supported by the
Alexander von Humboldt foundation. The work of SM was supported in part by Japan Society for the Promotion of Science Grants-in-Aid for Scientific Research No.~24K07017 and the World Premier International Research Center Initiative (WPI), MEXT, Japan. This work of KO is supported in part by Japan Society for the Promotion of Science (JSPS) KAKENHI Grant Numbers JP23KJ1162.
\end{acknowledgments}

\appendix

\section{Boundary conditions in an asymptotic region}
\label{app:boundary_conditions}
To determine the boundary conditions for the order-reduced equations with and without compensatory terms \eqref{eq:vaceqs_or}, \eqref{eq:or-phi}, \eqref{eq:or-psi}, \eqref{eq:or-phi_comp}, and \eqref{eq:or-psi_comp}, we solve the semiclassical Einstein equation and the scalar-field equations before the order-reduction in an asymptotic region.
In the asymptotic region, the metric function and the first $r$-derivatives of scalars are assumed to take the form of those in the Schwarzschild spacetime.
Hence, the first $r$-derivatives of scalars take the form of $\mathcal{O}(M/r^2)$ in the asymptotic region.
These suggest that, in this region, the metric function and the first $r$-derivatives of scalars can be expanded as
\begin{align}\label{eq:boundary-conditions-ansatz}
    f(r) &= 1 - \frac{2M}{r} + \frac{M^2 f_2 }{r^2} + \frac{M^3 f_3 }{r^3} + \frac{M^4 f_4}{r^4} + \frac{M^5 f_5}{r^5}+ \mathcal{O}\left( \frac{M^6}{r^{6}} \right)\,,
    \notag \\
    h(r) &= \left(1 - \frac{2M}{r} \right)^{-1} + \frac{M h_1}{r}+ \frac{M^2 h_2 }{r^2} + \frac{M^3 h_3 }{r^3} + \frac{M^4 h_4}{r^4} + \frac{M^5 h_5}{r^5}+ \mathcal{O}\left(\frac{M^6}{r^{6}} \right)\,,
    \notag \\
    \frac{\text{d} \varphi}{\text{d}r} &= \left. \frac{\text{d} \varphi}{\text{d}r} \right|_{\rm Sch} + \frac{M^2 d\varphi_3}{r^3} + \frac{M^3 d\varphi_4}{r^4} + \frac{M^4 d\varphi_5}{r^5} + \frac{M^5 d\varphi_6}{r^6} + \mathcal{O}\left( \frac{M^6}{r^{7}} \right)\,,
    \notag \\
    \frac{\text{d} \psi}{\text{d}r} &= \left. \frac{\text{d} \psi}{\text{d}r} \right|_{\rm Sch} + \frac{M^2 d\psi_3}{r^3} + \frac{M^3 d\psi_4}{r^4} + \frac{M^4 d\psi_5}{r^5} + \frac{M^5 d\psi_6}{r^6} + \mathcal{O}\left( \frac{M^6}{r^{7}} \right)\,,
\end{align}
where the first terms of the first $r$-derivatives of scalars is given in Eq. \eqref{eq:sol-phi-Sch}.
After solving the differential equations using the above expansions, we obtain the following expression:
\begin{align}\label{eq:boundary-conditions}
    f(r) &= 1 - \frac{2M}{r} - \frac{11638127 M^2 \hbar }{122472000 \pi r^4} +\frac{6194351 M^3 \hbar}{122472000 \pi r^5} + \mathcal{O}\left( \frac{M^6}{r^{6}} \right)\,,
    \notag \\
    h(r) &= \left(1 - \frac{2M}{r} \right)^{-1} + \frac{12998927  M^2 \hbar }{61236000 \pi r^4} +\frac{3253027 M^3 \hbar}{4898880 \pi r^5} + \mathcal{O}\left( \frac{M^6}{r^{6}} \right)\,,
    \notag \\
    \frac{\text{d} \varphi}{\text{d}r} &= \left. \frac{\text{d} \varphi}{\text{d}r} \right|_{\rm Sch} - \frac{2 M^2 \hbar}{135  \pi r^5} + \frac{20930800481 M^3 \hbar}{154314720000 \pi r^6} + \mathcal{O}\left( \frac{M^6}{r^{7}} \right)\,,
    \notag \\
    \frac{\text{d} \psi}{\text{d}r} &= \left. \frac{\text{d} \psi}{\text{d}r} \right|_{\rm Sch} + \frac{3915800293 M^3 \epsilon}{8641624320 \pi r^6} + \mathcal{O}\left( \frac{M^6}{r^{7}} \right)\,.
\end{align}
These asymptotics are consistent with the expectation that, in the asymptotic region, the source term in the Einstein equation falls off as $\mathcal{O}(M^4/r^6)$, so that the corresponding higher-order corrections to the metric functions begin at order $\mathcal{O}(M^4/r^4)$.
These corrections in the metric functions propagate to the scalar fields, and their effects first begin at the fifth or sixth order. The equations above can be used to calculate the error in a certain truncation.

\section{Backreaction under an alternative metric ansatz}
\label{app:alt_ansatz}
Here we examine whether our numerical integration breaks down due to the choice of the metric ansatz \eqref{eq:sssansatz}. To this end, in this section, we will analyze the backreaction under the metric ansatz:
\begin{equation}
    \text{d}s^2=-N(r)^2F(r)\text{d}t^2+\frac{\text{d}r^2}{F(r)}+r^2\text{d}\Omega^2\,.
\end{equation}
Although the $(t, r)$ coordinate chart becomes ill-defined at $F(r)=0$, in the static case one may move to the null coordinate $(v,r)$, which gives the same Einstein equations.
Under this metric ansatz, the function $N(r)^2$ is constant in the classical vacuum case; hence any deviation of $N(r)^2$ from its constant value characterizes the deviation from the Schwarzschild metric due to quantum vacuum fluctuations.
Moreover, the numerical evaluation of the deviation is stable in the present ansatz, whereas it is not under the metric ansatz \eqref{eq:sssansatz} because $f(r)$ and $h(r)$ are typically small and large, respectively.

\begin{figure}[htbp]
\includegraphics[width=0.55\linewidth]{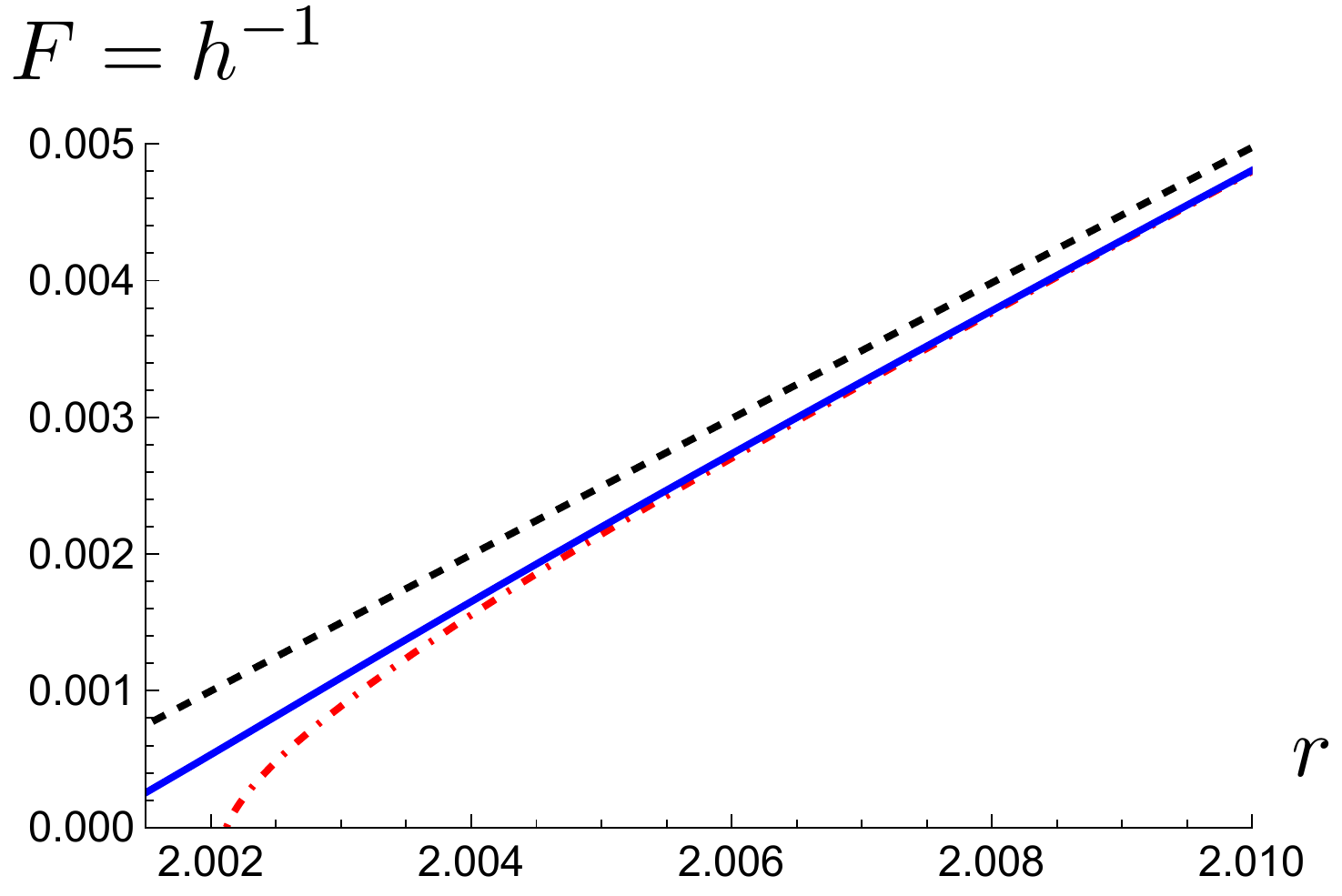}
\caption{Plot of the function $F(r)=h(r)^{-1}$ as a function of the radius. Different lines have the same meaning as in Fig.~\ref{fig:sol_f}. We use numerical values $M=1$, $\hbar=10^{-2}$, and $r_0=10^2$.
\label{fig:F_comparison_FNansatz}}
\end{figure}

\begin{figure}[htbp]
\includegraphics[width=0.55\linewidth]{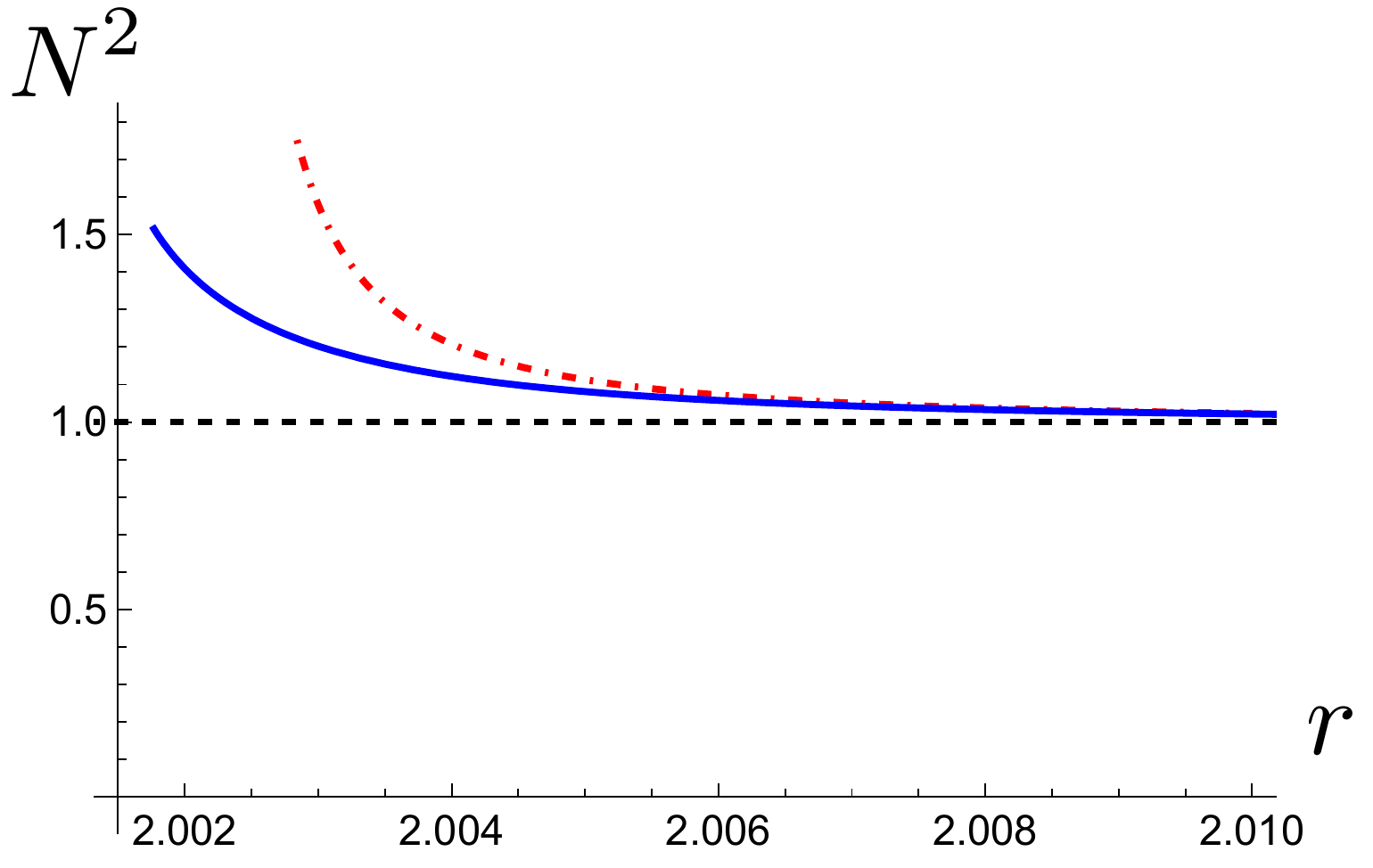}
\caption{Plot of the function $N(r)^2$ as a function of the radius. Different lines have the same meaning as in Fig.~\ref{fig:sol_f}.
\label{fig:N^2_comparison}}
\end{figure}

To perform the order-reduction procedure around the classical vacuum background, we find that all derivatives of the functions $F(r)$ and $N(r)$ are given by the following polynomials:
\begin{align}\label{eq:derrels_FN}
F^{(n)}&=(-1)^{n-1}\frac{n!(1-F)}{r^n}+\mathcal{O}(\hbar)\,,
\nonumber\\
N^{(n)}&=\mathcal{O}(\hbar)\,.
\end{align}
Using \eqref{eq:derrels_FN}, we apply the order-reduction to the RMV-RSET and the equations of motion for the scalar fields, and then integrate the semiclassical Einstein equations from the asymptotic region inward, with and without compensatory terms.
Regarding the boundary conditions for the metric functions, we use asymptotic solutions of $N(r)$ and $F(r)$ obtained by the transformations $N(r)^2=f(r)h(r)$ and $F(r)=h(r)^{-1}$, and the asymptotic solutions \eqref{eq:boundary-conditions}.
The results of the metric functions $F(r)$ and $N(r)$ are given in Figs. \ref{fig:F_comparison_FNansatz} and \ref{fig:N^2_comparison}, respectively. The Kretschmann scalar and the scalar-field profile exhibit behavior similar to that shown in Figs. \ref{fig:sol_dphi} and \ref{fig:curvature_inv}. From Fig. \ref{fig:N^2_comparison}, we find that the order-reduced quantum fluctuation with and without compensatory terms clearly becomes larger.
As for the radius at which the numerical integration breaks down, the computation reaches a slightly smaller radius than with the $(f, h)$ metric ansatz; the improvement is negligible.
These results suggest that, even though the effects of quantum fluctuations are more apparent, changing the metric ansatz does not lead to significant improvement.

\section{Behavior of the energy density on Killing horizons}\label{app:rhodiv}

Presenting a complete proof that the energy density for the order-reduced approximation with compensatory terms is unbounded on Killing horizons is out of the scope of this paper. However, it is possible to give a strong argument of plausibility that this is indeed the case by considering the following auxiliary spacetimes:
\begin{equation}\label{eq:metric_poly}
f(r)=\frac{r-r_{\rm h}}{r}\frac{P^{(2)}(r)}{r^2},\quad \frac{1}{h(r)}=\frac{r-r_{\rm h}}{r}\frac{Q^{(2)}(r)}{r^2}\,,
\end{equation}
where we have defined the second-order polynomials
\begin{equation}
P^{(2)}=r^2+p_1 r_{\rm h}r+p_2 r_{\rm h}^2,\quad Q^{(2)}=r^2+q_1r_{\rm h}r+q_2r_{\rm h}^2\,,
\end{equation}
and $\{p_1,p_2,q_1,q_2\}$ are constants. For $p_1=p_2=q_1=q_2=0$ we recover the Schwarzschild background, while other choices correspond to different spacetimes with Killing horizons.

\begin{figure}[htbp]
\includegraphics[width=0.6\linewidth]{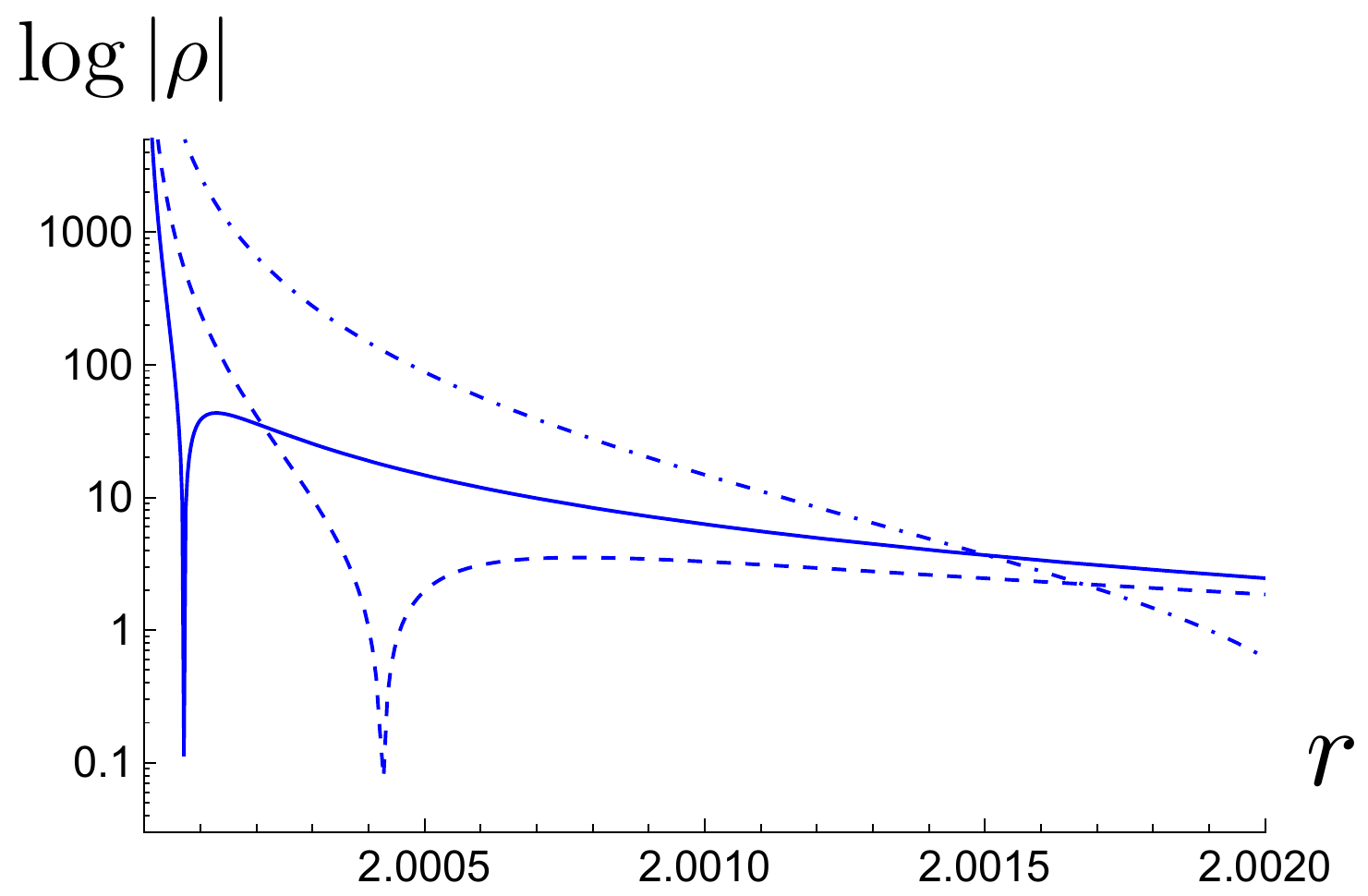}
\caption{
Logarithmic plots of the energy density obtained from order-reduced RMV-RSET with compensatory terms on background spacetimes described by \eqref{eq:metric_poly}. Here we set $(p_0, q_0)=(0.02, 0.03)$. The solid, dashed, and dot-dashed curves correspond to $(p_1, q_1)=(0.2, 0.1)$, $(p_1, q_1)=(1, 2)$, and $(p_1, q_1)=(-0.4, -0.5)$, respectively.
\label{fig:logRho_poly}}
\end{figure}

We have evaluated the corresponding expression of the order-reduced RMV-RSET on these spacetimes and observed that the energy density of the order-reduced RMV-RSET with compensatory terms is generically divergent no matter how the values of these coefficients are chosen (see Fig.~\ref{fig:logRho_poly}). Taking into account that there is no special feature in the family of spacetimes or the values of the coefficients considered, this strongly indicates that this divergence is a generic aspect of the approximation to the RSET with compensatory terms.

\providecommand{\href}[2]{#2}\begingroup\raggedright\endgroup

\end{document}